\documentclass{aa}  
\newcommand{\ergs}{${\rm erg \ cm^{-2} \ s^{-1}}$ }
\newcommand{\erg}{${\rm erg \ s^{-1}}$ }

\def\ltsima{$\; \buildrel < \over \sim \;$}
\def\simlt{\lower.5ex\hbox{\ltsima}}
\def\gtsima{$\; \buildrel > \over \sim \;$}
\def\simgt{\lower.5ex\hbox{\gtsima}}
\newcommand{\msun}{{\rm\,M$_\odot$}}

\newcommand{\srcs}{{\rm\,CSS1029}}
\newcommand{\src}{{\rm\,CSS1029 }}
\newcommand{\swift}{{\it\,Swift }}
\newcommand{\ppm}{{$\pm$}}

\usepackage{threeparttable}
\usepackage{graphicx}
\usepackage{txfonts}

\begin{document}

   \title{Discovery of late-time X-ray flare and anomalous emission line enhancement after the nuclear optical outburst 
in a narrow-line Seyfert 1 Galaxy}


   \author{W. J.~Zhang
          \inst{1}
          \and
	  X. W.~Shu
          \inst{1}
	  \and
	  Z. F.~Sheng
	  \inst{2}
	  \and
	  L. M.~Sun
	  \inst{1}
	  \and
	  L. M.~Dou
	  \inst{3}
	  \and
	  N. Jiang
         \inst{2}
	 \and
	 J. G.~Wang
	 \inst{4}
	  \and
	  X. Y.~Hu
	  \inst{2}
	  \and
	  Y. B.~Wang
	  \inst{2}
	  \and
	  T. G.~Wang
	  \inst{2}
          }

   \institute{Department of Physics, Anhui Normal University, Wuhu, Anhui, 241002, China\\
              \email{xwshu@mail.ahnu.edu.cn}
         \and
             CAS Key Laboratory for Researches in Galaxies and Cosmology, Department of Astronomy, University of Science and Technology of China, Hefei, Anhui 230026, China\\
             \email{shengzf@ustc.edu.cn; jnac@ustc.edu.cn}
         \and     
	     Department of Astronomy, Guangzhou University, Guangzhou 510006, China 
	  \and
	     Yunnan Observatories, Chinese Academy of Sciences, Kunming 650011, China 
	     }

 
  \abstract
   {
   CSS J102913+404220 is an atypical narrow line Seyfert 1 galaxy with an energetic optical outburst occurred in coincidence with its nucleus. 
   We present a detailed analysis of its multi-wavelength photometric and spectroscopic observations covering a period of decade since 
   outburst. 
   We detect {mid-infrared (MIR) flares delayed by about }two months relative to the optical outburst, with an extremely 
   high peak luminosity of $L_{4.6\mu m}>10^{44}$\erg. 
   The MIR peak luminosity is at least an order of magnitude higher than any known supernovae explosions, suggesting the optical outburst 
   might be due to a stellar tidal disruption event (TDE). 
   We find late-time X-ray brightening by a factor of $\simgt$30 with respect to 
   what is observed {about 100 days after the optical outburst peak},  
   followed by a flux fading by a factor of $\sim$4 within two weeks, 
   making it one of Active Galactic Nuclei (AGNs) with extreme variability.
   Despite the dramatic X-ray variability, there are no coincident strong flux variations in optical, UV and MIR bands.
   This unusual variability behavior has been seen in other highly accreting AGNs and 
   could be 
   {attributed to absorption variability.} 
	   In this scenario, the decrease in the covering factor {of absorber} with accretion rate could cause the X-ray brightening, possibly induced by the TDE. 
   Most strikingly, while the UV/optical continuum remains little changes with time, an evident enhancement 
   in the flux of H$\alpha$ broad emission line is observed, about a decade after the nuclear optical outburst,  
   which is an anomalous behavior never seen in any other AGNs.  
   Such an H$\alpha$ anomaly could be explained by the replenishment of gas clouds and excitation within Broad 
   Line Region (BLR) that originates, perhaps from the interaction of outflowing stellar debris with BLR. 
   The results highlight the importance of late-time evolution of TDE that could 
   affect the accreting properties of AGN, {as suggested by recent simulations.}
   }
   
   {}


\titlerunning{Late-time X-ray flare and anomalous emission line enhancement after the optical outburst in a NSy1}

     \maketitle
%

\section{Introduction} \label{sec:intro}

 It is believed that active galactic nuclei (AGNs) are powered by supermassive black holes 
 (SMBHs, $M_{\rm BH}\sim10^6-10^9~M_{\odot}$) accreting surrounding material. 
 The optical and UV photons are thought to originate from the accretion disk, 
 and a faction of disk emission transported into the hot corona, presumably in the 
 immediate vicinity of the SMBH, generating powerful X-rays at higher energies. 
 While rapid X-ray variability over a timescale from minutes to hours is ubiquitous in AGNs \citep{Ulrich1997, Gonz2012}, 
 large amplitude variations on longer timescales of years are not common, typically in the range $\sim50-200$\% \citep{Saxton2011, Yang2016, Middei2017, Boller2021}.  
 In addition, exotic AGN X-ray variability characterized by quasi-periodic modulations \citep{Gierli08, Song2020} or 
 even repeating short-lived X-ray eruptions \citep{Sun2013, Miniutti2019} are also found, providing a unique probe to the 
 accretion physics and X-ray radiation mechanisms. 

 Extreme X-ray variability by orders of magnitude on time scale of years 
 has been found in several types of AGNs \citep[e.g.,][]{Grupe2012a}, some of which are also accompanied by 
 the X-ray spectral changes, i.e., from Compton-thin to Compton-thick or vice versa. 
 This latter case can be naturally explained in terms of variations of absorption column 
 density along the line of sight \citep[e.g.,][]{Risaliti2005, Miniutti2014}. 
 Alternatively the dramatic X-ray flux and spectral variations can be ascribed to the 
 ``switch off and on" of the gas accretion onto SMBHs \citep{Guainazzi1998, Gilli2000}. 
 However, such a scenario is difficult to reconcile with the observations of 
 X-ray weak state in several highly accreting narrow line Seyfert 1s (NSy1s) and quasars, 
 where the UV flux remains almost constant despite the huge X-ray drop \citep{Gallo2011, Miniutti2012, Grupe2019}. 
 Since sources with high Eddington ratios such as NSy1s are expected to have a geometrically thick 
 accretion disk in innermost region, the disk self-shielding can explain the extreme 
 X-ray variability while UV/optical continuum at larger radii is not affected, 
 similar to the explanation proposed for the X-ray behavior of weak-line quasars \citep{Luo2015, Ni2018}. 
 Another scenario that has been invoked to account for the X-ray weakness is that the direct nuclear emission 
 is suppressed by the light-bending effect near the SMBH
 \citep{Miniutti2003, Miniutti2004}.

\begin{table*}
  \setlength{\tabcolsep}{1mm}{}
  \centering
    \caption{X-ray count rates, fluxes and optical magnitudes for \src observed by {\it Swift}.}
  \begin{tabular}{c c c c c c c c c c}  
    \hline 
    \hline 
Obs. date & {Exp. time} &Counts rate & X-ray flux & $M_{\rm UVW2}$ & $M_{\rm UVM2}$ & $M_{\rm UVW1}$ & $M_{\rm U}$ & $M_{\rm B}$ & $M_{\rm V}$ \\
    (1) & {(2)} & (3) & (4) & (5) & (6) &  (7) & (8) & (9)&(10) \\
    \hline  
 2010-04-06  &  {3490} &$0.31\pm0.10$ & $0.97\pm0.31$ & 18.06$\pm 0.04$ & 17.87$\pm 0.04$ &  17.56$\pm 0.04$ &  16.78$\pm 0.04$ &  16.35$\pm 0.04$ & 16.12$\pm 0.06$ \\
 2010-04-25  &{3010}& $0.12\pm0.07$ & $0.38\pm0.22$ & 18.38$\pm 0.05$ & 18.12$\pm 0.05$ &  17.71$\pm 0.05$ &  17.08$\pm 0.06$ &  16.61$\pm 0.06$ & 16.24$\pm 0.08$ \\
 2010-05-09 & {3285}&$0.11\pm0.07$ & $0.36\pm0.22$ & 18.50$\pm 0.06$ & 18.13$\pm 0.05$ &  17.92$\pm 0.06$ &  17.10$\pm 0.07$ &  16.75$\pm 0.07$ & 16.31$\pm 0.09$ \\
  2010-05-23 & {3985}&$<0.08^{\mathrm{a}}$ & $ <0.24^{\mathrm{a}}$ & 18.50$\pm 0.05$ & 18.35$\pm 0.06$ &  17.90$\pm 0.05$ &  17.10$\pm 0.06$ &  16.83$\pm 0.06$ & 16.59$\pm 0.09$ \\
 2015-03-26  & {175}& $<1.70$    & $<5.30$ & 19.44$\pm 0.20$ & 19.81$\pm 0.31$ &  18.96$\pm 0.21$ &  - &         -          &       -           \\
 2015-03-28  & {595}&$3.15\pm0.74$ & $9.52\pm2.24$ & 19.22$\pm 0.10$ & 19.41$\pm 0.14$ &  19.21$\pm 0.11$ &         -          &         -          &       -           \\
 2015-03-29 & {1451}&$2.86\pm0.45$ & $8.65\pm1.36$ & 19.39$\pm 0.18$ & 19.29$\pm 0.22$ &  19.34$\pm 0.19$ &         -          &         -          &       -           \\
 2015-04-02 & {284}&  $<1.05$    & $<3.27$ & 19.43$\pm 0.16$ & 19.47$\pm 0.20$ &  19.18$\pm 0.17$ &         -          &         -          &       -           \\
 2015-04-04 & {701} &$1.24\pm0.43$ & $3.75\pm1.30$ & 19.53$\pm 0.20$ & 19.14$\pm 0.20$ &  19.31$\pm 0.19$ &         -          &         -          &       -           \\
 2015-04-05 & {867}&$0.33\pm0.19$ & $0.98\pm0.60$ & 19.49$\pm 0.18$ & 19.64$\pm 0.25$ &  18.93$\pm 0.17$ &         -          &         -          &       -           \\
 2015-04-11 & {571}&$0.87\pm0.40$ & $2.63\pm1.22$ & 19.60$\pm 0.19$ & 19.91$\pm 0.29$ &  19.14$\pm 0.16$ &         -          &         -          &       -           \\
 2020-01-12 &{759}& $0.77\pm0.32$ & $2.25\pm0.91$ & 19.37$\pm0.13$ & 19.22$\pm0.11$ & 18.99$\pm0.15$     & 18.72$\pm$0.17 & 18.37$\pm$0.20  &  $>$18.15  \\
 2020-02-28 &{1650} &$<0.27$     & $<0.78$         & 19.17$\pm0.08$ & 19.27$\pm0.08$ & 19.03$\pm0.10$     & 18.50$\pm$0.11   & 18.73$\pm0.19$ & 17.79$\pm0.19$ \\
\hline
\hline
\end{tabular}
 \tablefoot{ Column (1): \swift XRT observing date; Column (2): XRT exposure time in unit of second; Column (3): Count rates in the 0.3--2 keV, in units of $10^{-2}$ cts s$^{-1}$; Column (4): X-ray flux in the 0.3--2 keV, in units of 10$^{-13}$ erg cm$^{-2}$s$^{-1}$;  Column (5): UVW2 ($\lambda_c=1928$\AA) magnitude in the AB system; Column (6): UVM2 ($\lambda_c=2246$\AA) magnitude in the AB system; Column (7): UVW1 ($\lambda_c=2600$\AA) magnitude in the AB system; 
 Column (8): U ($\lambda_c=3465$\AA) magnitude in the AB system; Column (9): B ($\lambda_c=4392$\AA) magnitude in the AB system;  Column (10): V ($\lambda_c=5468$\AA) magnitude in the AB system. $^\mathrm{a}$For all non-detections, the corresponding 3$\sigma$ upper limits are given.}

\end{table*}

 While extreme X-ray variations appear to be rare among standard AGNs, they are 
 more commonly seen in stellar tidal disruption events (TDE; see Komossa 2015 for a review).
 Such an event occurs when a star passes sufficiently close to a SMBH and is torn apart 
 when the tidal force of the SMBH exceeds its self-gravity. While roughly half of the 
 stellar material will be ejected, the other half will remain bound and eventually be accreted, 
 producing a luminous flare of electromagnetic radiation \citep{Rees1988}. 
 For a BH mass with $M_{\rm BH}\sim10^6M_{\odot}$, the peaking luminosity can reach $\sim10^{44}$\erg 
 and then declines by up to factors of $>$100 to the quiescent flux level of the host galaxy. 
 X-ray observations have revealed a population of TDE candidates with such large-amplitude X-ray variability 
 \citep[e.g.,][]{Komossa2015, Lin2017, Saxton2019}. 
 Note that similar extreme optical outbursts in the centers of galaxies have been detected recently, 
 but very few of them are associated with strong X-ray emission, the nature of which is still under debate 
 \citep{Kankare2017}. 
 In addition, theoretical works have suggested that TDEs can occur in AGNs and the event rate 
 may be even higher than in inactive galaxies \citep{Karas2007}. 
 Indeed, an increasing number of TDE candidates have been found in galaxies with pre-existing AGN activity 
 \citep[e.g.,][]{van2016, Blanchard2017, Shu2018, Liu2020}. 
 In this case, however, identifying whether nuclear outburst is due to a TDE or a particular episode of AGN variability 
 is not trivial \citep{Saxton2015, Auchettl2018, Trakhtenbrot2019, Hinkle2021}. 
 On the other hand, in the presence of TDE, the innermost regions of the accretion flow can be affected by the evolution 
 of stellar debris, leading to unique and remarkable AGN X-ray variabilities \citep{Blanchard2017, Chan2019, Ricci2020, Ricci2021}.  


 \begin{figure*}[htbp]
 \centering
 \includegraphics[scale=1.1]{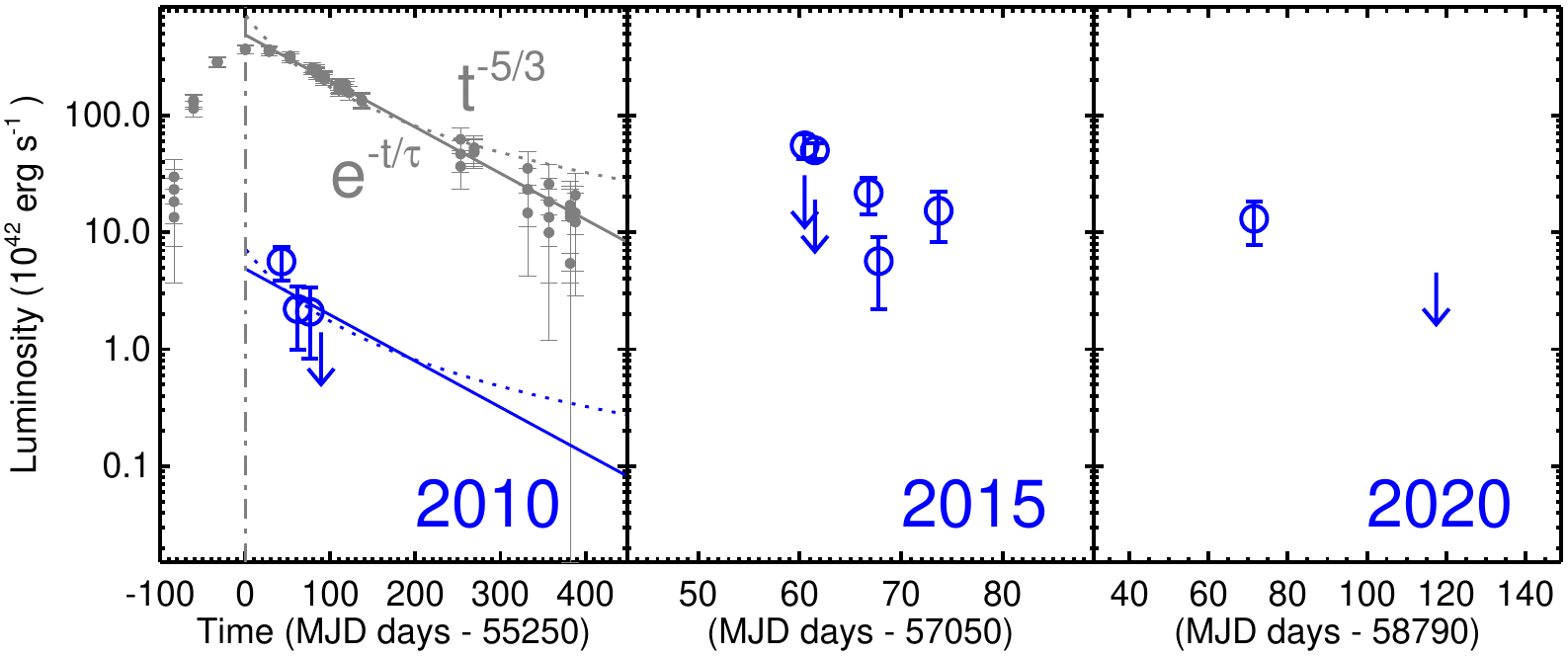}
 \label{xraylc}%
 \caption{
 X-ray light curves observed at different epochs. 
 In the left panel, the evolution of host-subtracted optical luminosity at the V-band is also shown (gray points, {in units of $10^{42}\rm erg~s^{-1}$)}. 
 The solid line shows the best exponential fit, while the dotted line 
 is the fit with the canonical $t^{-5/3}$ decay law. 
 The two best-fit lines are scaled downward by a factor of 100 to match the X-ray light curve. 
 }
 \end{figure*}

 CSSJ102912.58+404219.7 (CSS1029 hereafter) is a NSy1-like galaxy at $z=0.147$ that was discovered by the Catalina Real-time
 Transient Survey (CRTS; Drake et al. 2009) as an extraordinary optical transient in 
 the galaxy nucleus (Drake et al. 2011; D11 hereafter). Its extremely high peak 
 luminosity of $M_{\rm V}=-22.7$, distinctive features of hydrogen Balmer emission lines, and slow rise and fall of light curve over 
 a period of months led D11 to classify the transient as a ultra-luminous Type IIn supernovae. However, based on the similar spectral 
 properties to PS16dtm, another luminous transient occurred at the nucleus of a NSy1, \citet{Blanchard2017} 
 suggested that CSS1029 is more like a non-standard TDE \citep[see also][] {Kankare2017}. 
 The recent argument that the super luminous supernova (SLSN) ASASSN-15h is more consistent with a TDE \citep{Leloudas2016},  
 suggests that the bright optical outburst in CSS1029 may not necessarily be due to SLSN. 
 Here we present new optical spectroscopy as well as archival multi-wavelength observations of CSS1029, 
 and confirm the source as one of most luminous nuclear transients in terms of mid-infrared (MIR) luminosity, 
 making its association with Type IIn SLSN impossible. 
 In particular, we find a bright X-ray flare about five years since optical outburst, while no obvious 
 coordinated photometric variability at other bands is found. 
 Most strikingly, our spectroscopy follow-up observations about decade later reveal a significant 
 flux enhancement in the H$\alpha$ emission line, while the continuum emission remains almost constant. 
 This kind of H$\alpha$ variability is anomalous and rare among either TDEs or NSy1s. 
 In Section 2, we describe the observations and data reductions. 
 In Section 3, we present detailed analysis and results. 
 Possible explanations for the late time X-ray brightening and H$\alpha$ anomaly are given in Section 4. 

 \section{Observations and data}
 \par \src was first discovered as a bright optical transient on Feb 17 2010 (D11). 
  Prompt \swift observations of the source were performed four times in 2010. 
  There are also seven follow-up \swift observations performed on 2015, and two observations on 2020. 
  Details of the observation information were listed in Table 1. 
  Some results from the \swift 2010 observations have been presented in D11, while the data after 2015 have not yet been 
  fully reported. For the \swift/XRT (X-ray Telescope) data, all of them were operated in Photon Counting (PC) mode. 
 After reducing the data following standard procedures by \textit{xrtpipline}, we used {\tt Heasoft} (v6.19) to extract  
 spectrum with the task \textit{xselect}. The source spectrum was extracted in a circular region with a 40\arcsec~radius, 
 and we used another source-free circular region with a radius of 100\arcsec~for background. 
 We generated ancillary response function files (ARFs) by using \textit{xrtmkarf}, and the response matrix files (RMFs) were 
 downloaded with the latest calibration files.
 \par For \swift UVOT (the Ultra-violet Optical Telescope) data, all four observations in 2010 have data in six filters (UVW2, UVM2, UVW1, U, B, V). 
 By giving the source coordinate from optical position, we obtained magnitudes in the six bands for each observation 
 by using the task \textit{uvotsource}. 
 We performed flux extraction for source and background by using a circular region with 5\arcsec~radius and 
 a source-free region of circle with 50\arcsec~radius, respectively. 
 In observations performed in 2015, only three UV band filters have been used, and the source and background flux were extracted using 
 the same manner as for the 2010 data. 
 All the UVOT magnitudes are referred to the AB magnitude system, and listed in Table 1. 
 Note that the UVOT measurements are consistent with the results of D11, where only the data taken on 2010 are presented.

 We performed follow-up long-slit optical spectroscopy observations of \srcs, aiming to examine any spectral 
 evolution at later times. 
 The observations were taken twice with the Double Beam Spectrograph (DBSP) on the 200-in Hale telescope at Palomar Observatory (P200) 
 on Feb 2018 and Feb 2020. The latter one is quasi-simultaneously with the recent X-ray observations. 
 We used the D55 dichroic which splits the incoming photons into the 600/4000 (lines/mm) grating for the blue side, 
 and 316/7500 grating for the red side. 
 The grating angles were adjusted so that a nearly continuous wavelength coverage from 3300 to 10000 \AA~ is obtained, 
  with a spectral resolution of $R=\lambda/\Delta \lambda$$\sim$1080 at $\lambda$$\sim$6600\AA. 
 The data were reduced following the standard procedures. 
 The default slit width for the DBSP observations is 1.5\arcsec. 
 Although the observations have different seeing conditions, i.e., 3.3\arcsec~for Feb 2018 and 1.5\arcsec~for Feb 2020, 
 we used the same aperture size of $\sim$1.5\arcsec~to extract spectrum, with the aim to mitigate potential effect on 
 flux measurements of continuum and emission lines with different apertures. 
 The spectrum was then flux calibrated using standard star observations from the same night. 
 In addition, we also observed the source with the YFOSC (Yunnan Faint Object Spectrograph and Camera) 
 mounted on the Lijiang 2.4m telescope at the Yunnan Observatory 
 on April 2020, Jan 2021 and May 2021, respectively. For the former two YFOSC observations, we used 
 the spectral settings of G8 grating plus a slit width of 2.5\arcsec, while for latter a slit width 
 of 1.8\arcsec~was used. The observations have been performed under a typical seeing of $\sim$2\arcsec. 
 The G8 grating covers a wavelength range from 5100 to 9600\AA, with 
 a spectral resolution of $R$$\sim$380 at $\lambda$$\sim$6600\AA~for the 2.5\arcsec~slit, 
 and $R$$\sim$810 for the 1.8\arcsec~slit, respectively. 
 The YFOSC data were reduced using standard techniques in {\tt IRAF} and the same aperture of 1.5\arcsec~was used to extract spectrum. 
  Details on the spectroscopic observations, including date, seeing, exposure, spectral resolution and spectral extraction aperture 
 at each epoch are listed in Table 3. 

 \section{Analysis and results}

 \subsection{X-ray variability}
 As we mentioned earlier, D11 reported the X-ray detection from observations in 2010, but did not give the details of X-ray properties. 
 We will show that the source is clearly not detected in the last one of the four observations in 2010, indicating a significant flux decay. 
 Here we present a detailed and uniform analysis of all archival X-ray observations, especially for those that are not given in D11, 
 aiming to further investigate the X-ray flux and spectral evolution on longer timescales. 

 Figure 1 (left) plots the evolution of X-ray luminosity in the 0.3--2 keV for the observations taken on 2010, and  
 3$\sigma$ upper limit is shown for the non-detection\footnote{Since the 1$\sigma$ detection from the background fluctuation 
 results in less than one net count, we estimated conservatively the upper limit by requiring the detection of one net count, as did in \citet{Shu2020}.}. 
 The X-ray flux was converted from count rate using the best-fitting 
 powerlaw model to the stacked spectrum (Section 3.2).  
 It is evident that the source brightness declined by 
 {a factor of $\simgt$3 over a period of $\sim$50 days, from $5.6\pm1.8\times10^{42}$\erg to 
 the upper limit of $9\times10^{41}$\erg. 
 Such an evolution of X-ray brightness is not uncommon if due to a TDE \citep{Blanchard2017}, but no conclusive evidence can be drawn 
 as X-ray variability by even larger factors is often observed in NSy1s}. 
 To gain further insight into the nature of X-ray emission, we present the evolution of host-subtracted optical luminosity 
 at the V-band during the period of outburst (gray dots). 
 We estimate the host galaxy flux from the plateau of light curve before the optical outburst, 
 which is 17.51\ppm0.1 mag and consistent with that estimated by D11. 
 Note that we used the post-DR2 data from the CRTS survey to construct the host-subtracted light curve \citep{Graham2017}, 
 so it is slightly better sampled than that presented in D11 where only data up to $\sim$300 days after outburst are shown. 
 While the luminosity evolution can be well described by an exponential law $L=L_0e^{-(t-t_0)/\tau}$, 
 the fit with the popular $t^{-5/3}$ decay law is worse, especially for the data after $\sim$150 days since the peak.  
 Although the X-ray luminosity appears to follow the similar evolution (blue curves), it is nearly two orders of magnitude lower 
 than the optical luminosity at the same epochs. { 
 We will discuss possible reasons for the relative faintness of the X-ray emission 
 in Section 4.2. }
 


 {Although the source becomes invisible in the last observation performed on 2010,\swift detected it 
 again in 2015 in X-rays (Figure 1, middle). The highest luminosity recorded is $5.5\pm1.3\times10^{43}$\erg, 
 corresponding to a rebrightening by a factor of $>$30 relative to the previous non-detection. 
 In addition, the source declines in flux quickly by a factor of $\sim$4--10 over two weeks. 
 A previous study of \src by stacking individual observations claimed the detection of X-ray emission in 2015 as the same order of magnitude 
 as the first observation in 2010 \citep{Auchettl2017}. However, this is not true when considering the flux decay and particularly the 
 difference between the lowest state in 2010 and highest state in 2015, which are not considered in past analyses. 
 While the source is detected in the first observation performed on Jan 2020 with $\sim6\pm2$ net counts, 
  it becomes invisible again in the second observation about a month later (Figure 1, right). 
 Although we lack enough X-ray observations to monitor the luminosity evolution since 2015, the above analysis suggests 
 that the long-term X-ray variability is presented in \src, by factors of $\sim$10 on time-scales as short as a few tens of days, 
 consistent with that seen in NSy1s.}

 %


 \begin{figure}[htbp]
 \centering
 \rotatebox{-90}{\includegraphics[scale=0.3]{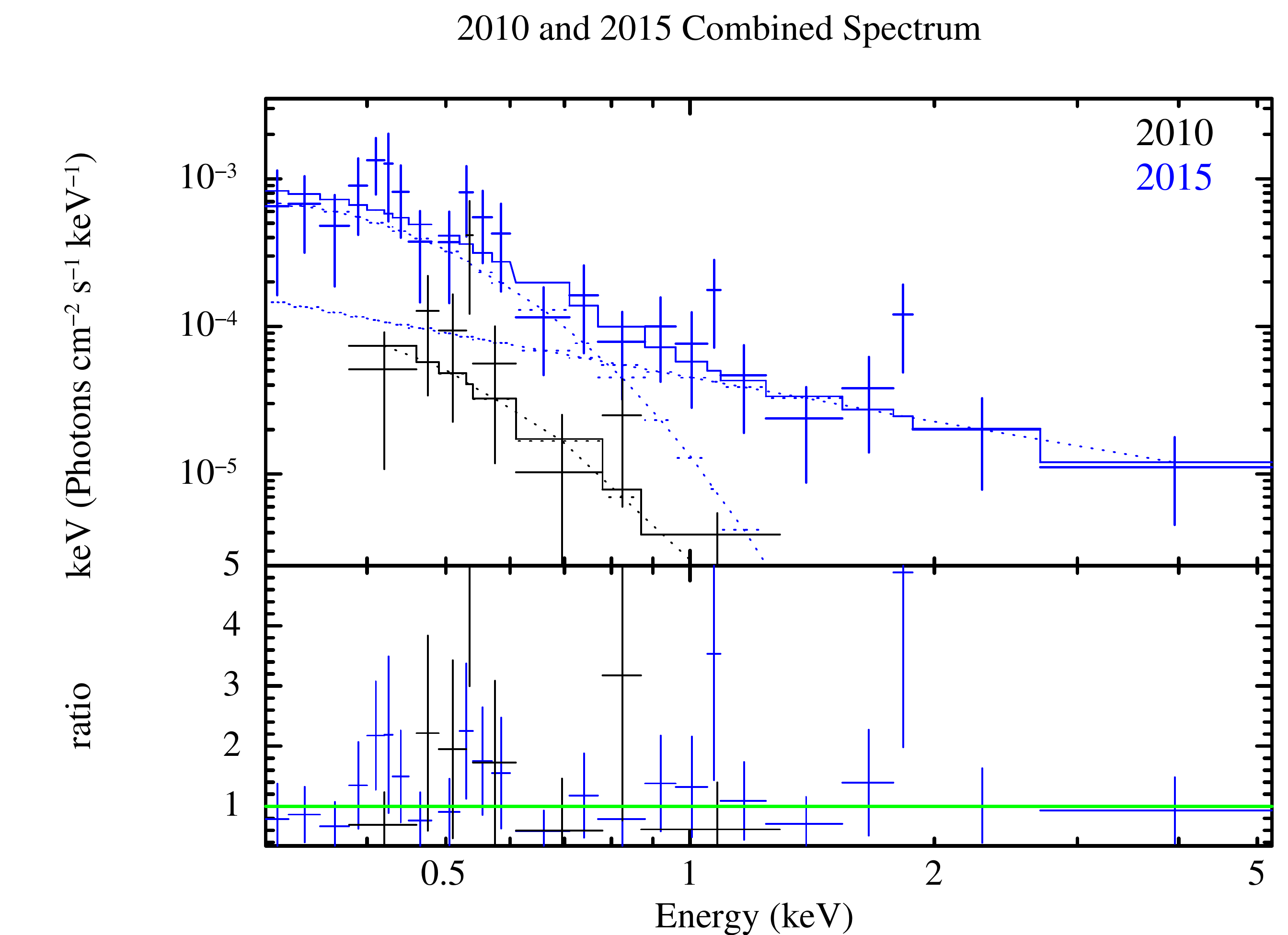}}
 \label{xspec}%
 \caption{The X-ray spectrum by stacking three observations in 2010 (black) and five observations in 2015 (blue). 
 {The dotted line shows the best-fitted model consisting of a powerlaw and a blackbody component. 
  Note that the spectral fittings were performed jointly to the 2010 and 2015 data based on this model}. 
 The corresponding data to model ratios are shown in bottom panel.
 }
 \end{figure}

 \subsection{X-ray spectral analysis}
 Because of short exposures, the spectral signal-to-noise ratios (S/Ns) for most of the individual \swift observations are not sufficient
 to perform meaningful fittings. Thus, in order to obtain a spectrum with better S/N, we combined the spectra from individual observations using the 
 FTOOLS task {\tt addascaspec}. We added spectral files separately from the three \swift observations taken in 2010 (\swift 2010), 
 and five observations in 2015 (\swift 2015). Non-detections are excluded in the spectral stacking (Table 1). 
 We obtained $17\pm5$ and $82\pm9$ background subtracted counts in the 0.3--5 keV for the \swift 2010 and 2015, respectively. 
 Since only $6\pm3$ net counts are detected in the first \swift observation performed on 2020, we do not consider the 2020 data in the following spectral analysis. 
 The \swift 2010 spectrum appears soft, with most of the counts falling below 1.5 keV ($\sim$80\%). 
 For the \swift 2015 observations, we detected 11 net counts at energies above 1.5 keV, indicating the presence of harder X-ray emission (Figure 2). 
 In the following spectral analysis, we grouped the data to have at least 2 counts in each bin to ensure the use of $C-$statistic for the spectral fittings. 
 The Galactic column density was considered and fixed at $N_{\rm H}^{\rm Gal}=1.23\times10^{20}$ cm$^{-2}$. 
 {All statistical errors given hereafter correspond to $\Delta C=$2.706 for one interesting parameter}, unless stated otherwise.

\begin{table}
  \centering
  \renewcommand\arraystretch{1.3}
   \setlength{\tabcolsep}{1mm}{
     \caption{Spectral fitting results for the \swift 2010 and 2015 data. }
  \begin{tabular}{c c c c c}
    \hline 
    \hline 
    Model & $N_{\rm H}$  & $\Gamma$   & $kT$ & {\bf $C$}/dof \\
          & ($10^{21}$ cm$^{-2}$) & & (eV) & \\
    \hline  
    & & {\it Swift} 2010 & & \\
    \hline  
 powerlaw &  $<5.6$ &  $3.89^{+2.75}_{-1.94}$ & & 7.1/7 \\
 blackbody &  0.12$^f$ &  & $126^{+90}_{-57}$  & 7.2/7 \\
 \hline
 & & {\it Swift} 2015& & \\
  \hline
 powerlaw &  $<1.0$  &  $3.36^{+0.38}_{-0.36}$ &   &  36.4/40 \\
 powerlaw+bbody & 0.12$^f$ & $2.16^{+1.02}_{-1.22}$ & $101\pm28$ & 33.3/39 \\
\hline
\hline

\end{tabular}}
  \tablefoot{ $^f$The parameter is fixed in the spectral fittings.}
  
\end{table}

 \begin{figure*}[htbp]
 \centering
 \includegraphics[scale=0.57]{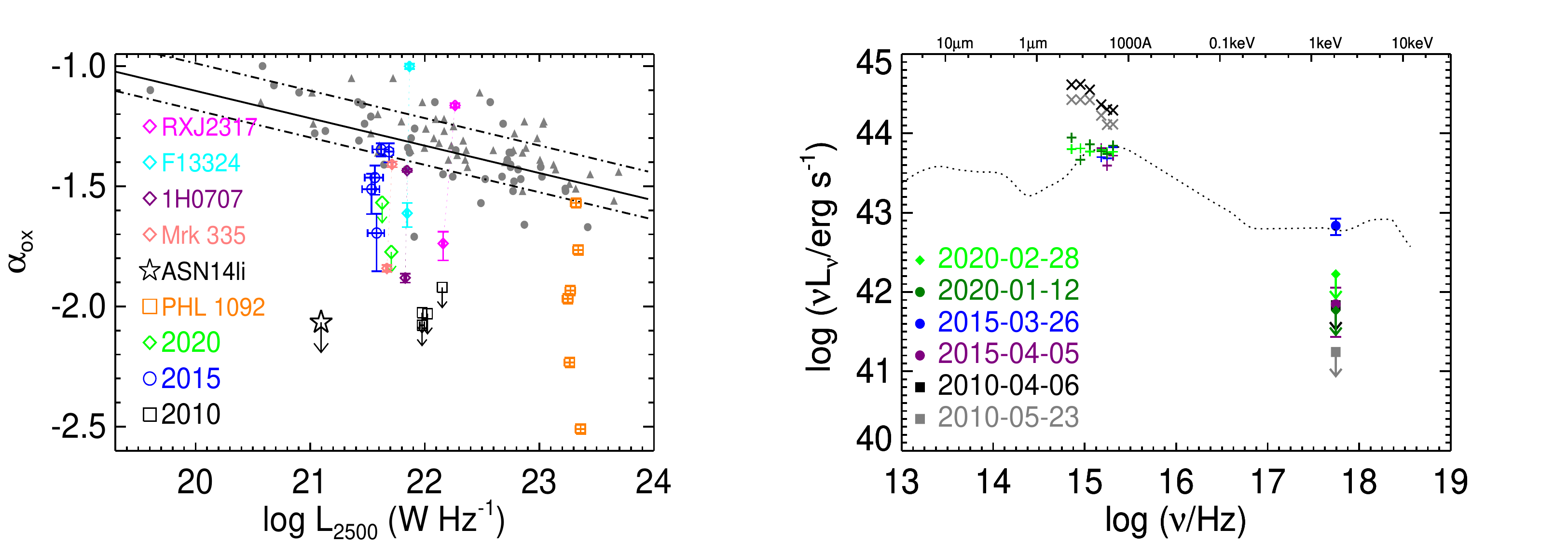}
 \label{sedaox}%
 \caption{
 { {\it Left:} UV-to-X-ray spectral index $\alpha_{\rm ox}$ versus 2500\AA~luminosity for \src (black, green and blue symbols). 
 The solid line represents the fit to the X-ray selected AGNs from \citet{Grupe2010}. The NSy1s are shown in gray dots, 
 while the broad-line Sy1s are shown in gray triangles. For comparison, we also plot the $\alpha_{\rm ox}$ for four typical 
 NSy1s with extreme X-ray variations. The $\alpha_{\rm ox}$ were calculated in the same manner as \srcs. 
 For clarity, only the two data points representing observations between the high and low sate 
 are shown. 
 The open star represents the data for the thermal TDE ASASSN-14li in the flare state \citep{van2016}. }
 {\it Right:} Optical to X-ray SED of \srcs. For clarity, only the data for the highest and lowest state within each epoch are shown. 
 The median SED from radio quiet quasars (Elvis et al. 1994) is also shown for comparison (dotted line), for which the luminosity  
 is scaled to that of UVW2 obtained in 2015 high state.  
 }
 \end{figure*}

 As a first step, we performed the spectral fittings to the combined \swift 2015 spectrum which has the best spectral quality, using 
 a simple absorbed powerlaw model. The fitting result is acceptable, with a 
 powerlaw index $\Gamma=3.36^{+0.36}_{-0.34}$ and $C/dof=36.4/40$. 
 The intrinsic absorption is not required, with an upper limit on the column density of $N_{\rm H}<1\times10^{21}$ cm$^{-2}$. 
 Although loosely constrained, the power-law component is relatively steep among AGNs \citep{Grupe2010}. 
 Thus we attempted to fit the spectrum with 
 a blackbody ({\tt zbbody} in {\tt XSPEC}), but found an excess of emission at energies above $\sim$1.5 keV. The addition of a powerlaw to 
 the model improves the fit significantly ($C$-value is decreased by 24.9 for two extra parameters). In this case, we obtain a blackbody 
 temperature of $kT_{\rm BB}=101\pm28$ eV, and the powerlaw component becomes flatter with $\Gamma=2.16^{+1.02}_{-1.22}$, which are  
 consistent with the typical values observed in NSy1s.
 
 Similar spectral analysis was also performed for the \swift 2010 data, which can be described by a simple absorbed powerlaw model with 
 photon index $\Gamma=3.89^{+2.75}_{-1.94}$. 
 The photon index is comparable to that observed in 
 TDEs which can be ascribed to the thermal disk emission at soft X-ray bands \citep{Komossa2015}.
 We found a single blackbody model providing an equally good fit to the \swift 2010 data, without 
 requirement for an additional powerlaw component. 
 Since the spectral quality is poor for the 2010 data, we cannot tell whether the source 
 is intrinsically weak at hard X-rays or such a component is simply eluded detection due to insufficient exposure. 
 Given that the 0.3--1 keV spectral shape is consistent with each other between the 2010 and 2015 observations, we 
 then considered a joint fit to the two data sets in the 0.3--5 keV. We used the best-fitting model for the 2015 spectrum, comprising 
 a thermal blackbody model plus a power law, modified by the Galactic absorption. 
 All the parameters are fixed to be the same except for flux normalizations.
 We obtained an acceptable joint fit ($C/dof=43.2/48$) with $kT\sim101$ eV and $\Gamma\sim2.04$. 
 Based on this model, we found an 0.3--2 keV flux of $7.02\times10^{-14}$\ergs and $5.42\times10^{-13}$\ergs, 
 and 2--10 keV flux of $1.98\times10^{-14}$\ergs and $1.53\times10^{-13}$\ergs for the 2010 and 2015 data, respectively. 
 Note that the hard X-rays are clearly not detected for the 2010 observations, hence the above 2--10 keV flux from the model 
 extrapolation would serve as only an upper limit.

 \subsection{UV emission and optical-to-X-ray spectral energy distribution}
 \par 
 {As we shown above, the X-ray emission appears relatively faint during the period of optical outburst in \srcs. 
 In order to gain insights into the X-ray properties, we turn now to the analysis of the 
 UV emission and its flux ratio to X-rays. }
 All the \swift observations provide simultaneous UV photometric data through UVOT in the filter UVW2 ($\lambda_c=1928$\AA), 
 UVM2 ($\lambda_c=2246$\AA) and UVW1 ($\lambda_c=2600$\AA), respectively. 
 For the observations performed on 2010 and 2020, the simultaneous optical data are also available (Table 1). 
 It is interesting to note that while the X-ray flux increases by a factor of up to $>$30 between 2010 and 2015 observations, 
 the UV variations, however, have much smaller amplitudes.  
 For example, the maximum UV variation for the UVW2 is $\sim$1.54 mag, corresponding to only a factor of $\sim$4 drop  
 in the monochromatic flux between the two epochs. 
 The UV variations are neither strongly correlated with the X-ray ones within each epoch.

 The availability of simultaneous UV and X-ray data allows us to derive reliably the UV-to-X-ray slope for each 
 observation. 
 The UV-to-X-ray slope is often defined by monochromatic flux ratio between 2 keV and 2500\AA~in the rest frame,
 parameterized with $\alpha_{\rm ox}=0.384$log$(f_{\rm 2keV}/f_{\rm 2500\AA})$. 
 To calculate the 2500\AA~flux, we used the observed flux at the bluest UV band ($\sim$1928\AA, UVW2), 
 which is expected less affected by host contamination. 
 Then the UVW2 flux was extrapolated to the 2500\AA~by assuming a typical UV spectral index of $\alpha=0.65$~($S_{\nu}\propto\nu^{-\alpha}$, 
 Grupe et al. 2010). 
 We note that adopting a different slope of $\alpha$ between $1$ and 0 in the extrapolation would 
 result a change by $\simlt50$\% in the 2500\AA~flux and hence by $\simlt20$\% in the $\alpha_{\rm ox}$. 
 {The X-ray flux density at 2 keV was estimated by the best-fitting model that was used to 
	 perform the joint fit to the 2010 and 2015 data, and corrected for the Galactic absorption (Section 3.2). 
	 The model consists of a powerlaw with $\Gamma=2.04$ and a blackbody component with $kT=101$ eV.  
 In order to avoid introducing model-dependent biases, we adopted the same model  
 to estimate the 2 keV flux for the {\it Swift}2020 data.} 
 Since no X-rays above 1.5 keV are detected for the 2010 and 2020 observations, we considered all the $\alpha_{\rm ox}$ 
 as upper limits for these epochs. 
 Our measurements imply that \src has a much steeper UV to X-ray spectral slope ($\alpha_{\rm ox}\simlt-2$) 
 in 2010 than that in 2015 ($\alpha_{\rm ox}\simeq-1.5$). 

 It is well known that the $\alpha_{\rm ox}$ are correlated with the 2500\AA~luminosity for AGNs, which are 
 increasingly X-ray weak for higher UV luminosities \citep[e.g.,][]{Steffen2006}. 
 Such a relation for the X-ray selected AGNs is shown by the solid line \citep{Grupe2010} 
 in Figure 3 (left), 
 and the dotted line represents the approximately 1$\sigma$ scatter of the distribution. 
 The NSy1s from the sample are shown in gray filled circles, while the broad-line Sy1s are shown in triangles. 
 It can be seen that the $\alpha_{\rm ox}\simeq-1.35$ at the highest state (rebrightening stage in 2015) 
 is quiet typical for \src as a normal NSy1. 
 However, the source became increasingly X-ray weak in the subsequent observations, down to an 
 X-ray level comparable to that obtained with the most recent observations in 2020, falling 
 significantly below the extrapolation of the $\alpha_{\rm ox}-L_{\rm 2500\AA}$ relation. 
 The average $\Delta\alpha_{\rm ox}=\alpha_{\rm ox}-\alpha_{\rm ox,exp}=-0.191$ and $<$$-$0.379 
 for the 2015 and 2020 data\footnote{$\alpha_{\rm ox,exp}$ is the $\alpha_{\rm ox}$ expected 
 from the Grupe et al. (2010) $\alpha_{\rm ox}-L_{\rm 2500\AA}$ relation.}, 
 corresponds to be X-ray weaker than $\alpha_{\rm ox,exp}$ at a 2.38$\sigma$ and 4.74$\sigma$ level, 
 respectively. 
 In comparison with the 2015 and 2020 data, \src is even more (hard) X-ray weak in 2010 according to the $\Delta\alpha_{\rm ox}$. 
  Such an extreme X-ray weakness is unusual among AGNs, when compared with the NSy1s in an extreme X-ray faint state, such as 
 Mrk 335 \citep{Grupe2012b}, 1H0707 \citep{Fabian2012}, IRAS F13324-3809 \citep{Buisson2018} and RX J2317.8-4422 \citep{Grupe2019}. 
 {This is similar to the X-ray weak quasar phenomenon, such as PHL 1092 from {\it XMM-Newton} monitoring observations \citep{Miniutti2012}, 
 suggesting \src might be one of AGNs with extreme $\alpha_{\rm ox}$ variability ever found. 
 Alternatively, such an X-ray weakness might be 
 expected if the X-ray emission was dominated by a soft blackbody component }
 as originated from the transient disk accretion in the process of TDE. 
 In fact, the ultrasoft X-ray spectra have been found common in TDEs \citep{Komossa2015, Auchettl2017}. 
 We also plot the $\alpha_{\rm ox}$ (open star in Figure 3 (left)) for the best-studied thermal TDE ASASSN-14li 
 in the flare state \citep[e.g.,][]{van2016}, and find it is comparable to the 2010 data for \srcs, 
 suggesting a possible connection between the two. 


 A further perspective on the unusual X-ray properties of \src comes from the detailed comparison of 
 optical-to-X-ray spectral energy distribution (SED) between different epochs, as shown in Figure 3 (right). 
 For clarity, we show only the data from the high and low state for the three epochs. 
 The recent \swift observations in 2020 revealed a similar flux as the one in the last observation of 2015, 
 followed by a drop by a factor of $\simgt$3 within $\sim$50 days, indicating a possibly new X-ray variability.  
 For comparison, the median SED of radio-quiet quasars, scaled to the UVW2 flux in the 2015 high state, 
 is also shown (dotted line). 
 The UV-to-X-ray SED of \src in the 2015 high state is consistent with those of typical quasars, 
 while it is X-ray weak by an order of magnitude in the low state, likewise for the 2020 data. 
{We note that the optical-to-X-ray SEDs for observations in 2010 might be 
  different from the median SED of typical quasars, even for the high state data, 
  suggestive of extreme X-ray weakness of intrinsic AGN emission \citep[e.g.,][]{Miniutti2012, Liu2019}. 
  The results are consistent with the above analysis of $\alpha_{\rm ox}$ variability. 
  }

 \begin{figure}[ht]
 \centering
 \includegraphics[scale=0.33]{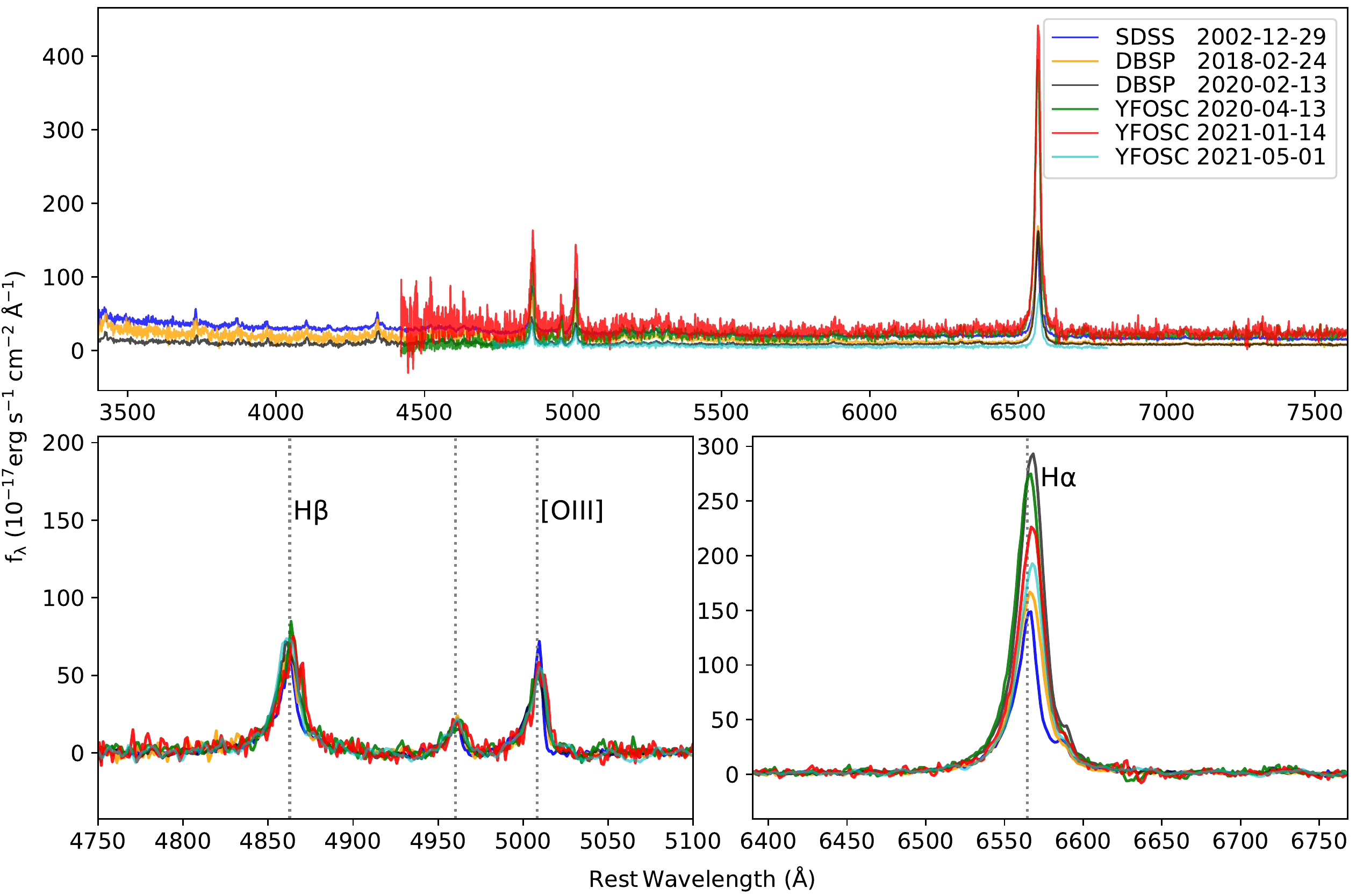}
 \caption{
 The P200/DBSP spectrum observed on Feb 2018 (orange) and Feb 2020 (gray), YFOSC spectrum observed on April 2020 (green), 
 Jan 2021 (red) and May 2021 (cyan), in comparison with the archival SDSS spectrum taken on 2002 (blue). 
 The lower panel shows a zoomed-in view of the starlight- and continuum-subtracted spectra in the $H_{\beta}$ and $H_{\alpha}$ region, 
 which are calibrated using the {\sc [O iii]} narrow-line flux. }
 \label{optspec}
 \end{figure}

 \subsection{Optical spectrum and its evolution}

\begin{table*}
     \caption{Spectroscopic observations and emission line fitting results. }
   \centering
  \begin{tabular}{c c c c c c c}
    \hline 
    \hline 
Obs. log & SDSS & P200 & P200 & YFOSC & YFOSC & YFOSC \\
    \hline  
Obs. date & 2002 Dec & 2018 Feb & 2020 Feb & 2020 April & 2021 Jan &  2021 May \\
Seeing (arcsec)   &   1.78-2.2         & 3.3 & 1.5 &  2.1   &  1.7        & 2.1 \\
Aperture width (arcsec) & 3$^\dag$        & 1.5 & 1.5 & 1.5 &  1.5 & 1.5 \\
Spectral resolution ($\lambda/\Delta\lambda$)$^\ddag$ & 1780  & 1080  & 1080 & 380 & 380 & 810\\
Exposure (second) &   3000          & 2700  &  800  &  1800  &  1800 & 2400 \\
 \hline
Parameters \\
 \hline
 H$_{\alpha \rm b}$ ($10^{-14}$erg cm$^{-2}$s$^{-1}$) &  $0.85\pm0.60$ & $1.25\pm0.1$ & $2.8\pm0.09$ & $2.6\pm0.08$ &  $2.0\pm0.23$ & $1.7\pm0.01$\\
 H$_{\alpha \rm m}$ ($10^{-14}$erg cm$^{-2}$s$^{-1}$) &  $1.42\pm0.07$ &$2.76\pm0.1$& $4.1\pm0.1$& $4.1\pm0.09$ & $3.4\pm0.27$ & $2.9\pm0.01$ \\
H$_{\alpha \rm n}$ ($10^{-14}$erg cm$^{-2}$s$^{-1}$) &  $0.39\pm0.04$ & $0.39\pm0.01$& $0.39\pm0.01$ & $0.39\pm0.01$ &  $0.39\pm0.01$ & $ 0.39\pm0.01$ \\
 FWHM [H$_{\alpha \rm b}$] (km s$^{-1}$) & $2860\pm170$ & $3380\pm512$ & $2653\pm111$ & $2790\pm92$ &  $3046\pm236$ & $3476\pm32$\\
 FWHM [H$_{\alpha \rm m}$] (km s$^{-1}$) & $807\pm38$ & $1034\pm32$ & $846\pm13$ & $892\pm11$ &  $911\pm21$ & $ 920\pm2.5$\\
 FWHM [H$_{\alpha \rm n}$] (km s$^{-1}$) & $318\pm15$& 318$^{f}$ & 318$^{f}$ &318$^{f}$ &  318$^{ f}$ & 318$^{f}$\\
 H$_{\beta \rm b}$ ($10^{-15}$erg cm$^{-2}$s$^{-1}$) &  $6.7\pm0.3$ & $7.8\pm1.6 $& $7.5\pm0.9$ & $7.4\pm0.6$ & $7.9\pm2.0$ &  $7.4\pm0.2$\\
 H$_{\beta \rm m}$ ($10^{-15}$erg cm$^{-2}$s$^{-1}$) &  $3.2\pm0.4$ & $4.9\pm1.6$ & $5.6\pm0.9$ & $6.3\pm0.5$ &  $5.5\pm2.2$ &  $5.5\pm0.3$\\
 H$_{\beta \rm n}$ ($10^{-15}$erg cm$^{-2}$s$^{-1}$) &  $1.3\pm0.3$ & $1.3\pm0.1$ & $1.3\pm0.08$ & $1.3\pm0.01$ &  $1.3\pm0.01$ &  $1.3\pm0.01$ \\
 FWHM [H$_{\beta \rm b}$] (km s$^{-1}$) & $2721\pm154$ & $2821\pm625$ & $2449\pm213$ & $2787\pm36$ &  $2562\pm222$ & $2099\pm87$ \\
 FWHM [H$_{\beta \rm m}$] (km s$^{-1}$) & $736\pm67$ & $746\pm88$ & $920\pm86$ &$860\pm62$ &  $866\pm229$ & $650\pm24$\\
 FWHM [H$_{\beta \rm n}$] (km s$^{-1}$) & $318\pm15$ & 318$^{f}$ & 318$^{ f}$&318$^{f}$& { 318$^{f}$} & 318$^{f}$\\
 $L_{5100\AA}$ ($10^{43}$erg s$^{-1}$) & $6.6\pm1.0$ & $4.2\pm1.0$ & $4.05\pm1.0$ & $3.7\pm1.0$ &  $3.8\pm1.0$ &  $4.9\pm1.0$\\
\hline
\hline
\end{tabular}
\tablefoot{$^{\dag}$  It corresponds to the fiber size to extract the SDSS spectroscopy, which has a diameter of 3\arcsec. 
 $^{\ddag}$ The spectral resolution at $\lambda=6600$\AA. 
 $^{f}$The parameter is fixed at the best-fit value from the spectral decomposition 
 of the pre-flare SDSS data. }
\end{table*}

 High-amplitude X-ray flaring has been detected in multiple types of AGNs, some of which are 
 accompanied by a strong change in the flux of the broad Balmer lines \citep{Shappee2014, Parker2016, 
 Oknyansky2019}, probably due to an increase in the accretion rate. 
 We performed two follow-up spectroscopy observations with P200 on Feb 2018 (P18) and Feb 2020 (P20), 
  and three with YFOSC on April 2020 (Y20), Jan 2021 (Y21a) and May 2021 (Y21b), respectively. 
 The 2020 observations were carried out promptly after the recent detection of X-rays. 
 Figure \ref{optspec} (upper panel) shows the follow-up spectra as well as the earlier spectrum 
 from the Sloan Digital Sky Survey (SDSS) for comparison. 
 It appears that spectral variations are present between different epochs. 
 To account for different observing conditions and apertures that may lead to spectral differences, 
 we calibrate the spectra using the observed {\sc [O iii]}$\lambda$5007 flux, under the assumption that 
 the {\sc [O iii]} narrow-line flux does not vary over the timescale of interest \citep[e.g.,][]{Peterson2013}. 
 The calibrated, starlight and continuum-subtracted spectra in the $H_{\alpha}$ and $H_{\beta}$ region 
 are shown in Figure \ref{optspec} (lower panel). 
 Significant changes in the $H_{\alpha}$ emission line is obvious, while the same trend is not visible 
 in the $H_{\beta}$ line.  
   
  \begin{figure*}[ht]
  \centering
  \includegraphics[scale=0.53]{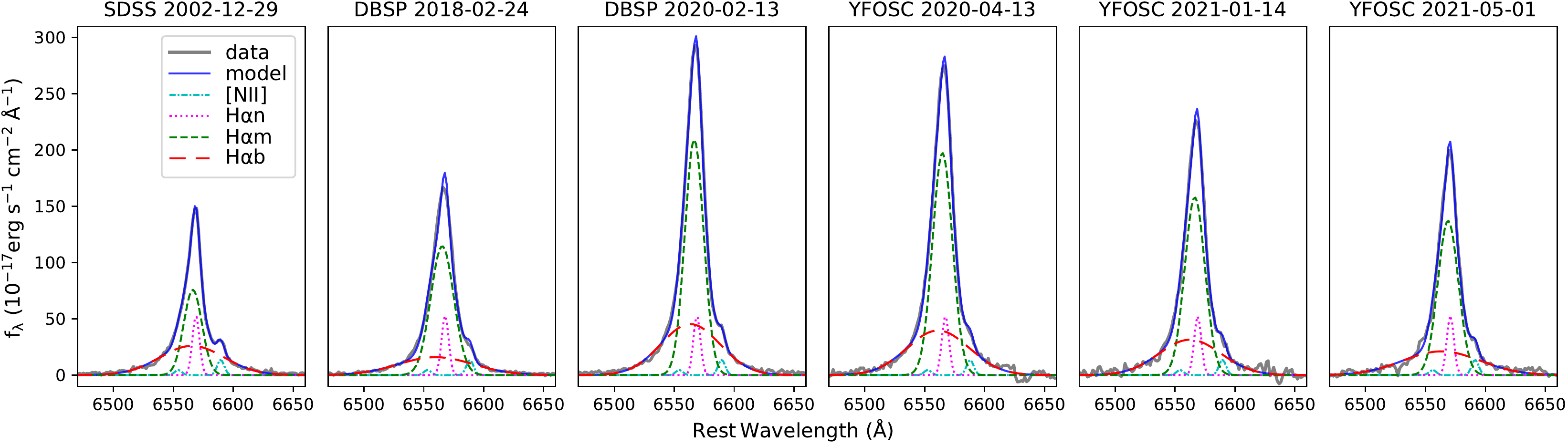}
  \caption{ Illustration of the emission line profile fitting in the $H_{\alpha}$+{\sc [N ii]} region observed at different epochs.}
  \label{haprofile}%
  \end{figure*}
   
  To further explore the $H_{\alpha}$ profile changes, we performed detailed spectral fittings to measure 
  the AGN continuum and the emission lines. 
   Firstly, the Galactic extinction was dereddened using the dust map provided by \cite{Schlegel1998}.
  Then we tried to decompose the spectra into a host galaxy and AGN component with the principal component analysis method 
  \citep{Yip2004}. However, we obtained negative flux for the host galaxy, which means the spectra are entirely dominated by the AGN emission.
  Therefore, we did not consider the host galaxy component in our following spectral analysis. 
  The continuum was then modeled by a power law plus a third-order polynomial with the region around the broad emission lines masked out, 
  while the optical and UV Fe {\sc ii} were modeled by empirical templates from literature \citep{Boroson1992, Vestergaard2001}. 
  For the Balmer emission lines, \citet{Drake2011} found that there are three significant components with narrow, 
  medium, and broad velocity widths. We hence used three Gaussians to model the  
  $\rm H\alpha$ (and $\rm H\beta$) line.  For {\sc [O iii]} doublets, each narrow line is fitted by two Gaussians, 
  while that of {\sc [N ii]} and {\sc [S ii]} doublets is represented by a single Gaussian. During spectral fittings, 
  the profiles and redshifts of narrow lines are tied. In addition, flux ratios of the {\sc [O iii]} doublets 
  $\lambda$5007, $\lambda$4959 and {\sc [N ii]} $\lambda$6583, $\lambda$6548 are fixed to their theoretical values. 
  Considering the lower S/N in the $\rm H\beta$ region of the P18 and Y20 spectrum, the upper limit on the FWHM 
  (Full Width at Half Maximum) of the $\rm H\beta$ broad component is bounded and set by that of $\rm H\alpha$. 
  The narrow line components of Balmer lines are assumed to be constant and fixed to that of preflare SDSS spectrum. 
   The median flux between rest-frame 5095 and 5105 \AA\, is taken to calculate the 5100 \AA\,luminosity.
 Uncertainties on the spectral parameters, including the 5100\AA~luminosity, were derived using Monte Carlo simulations, 
 where the observed spectrum was randomly perturbed by adding Gaussian noise at each wavelength bin, with amplitude 
 determined by the flux error. 
 This was repeated 100 times, resulting in a distribution of the best-fitting parameter of interest, 
 the standard deviation of which was considered as its 1$\sigma$ error.

   

 \begin{figure}[htbp]
 \centering
 \includegraphics[scale=0.53]{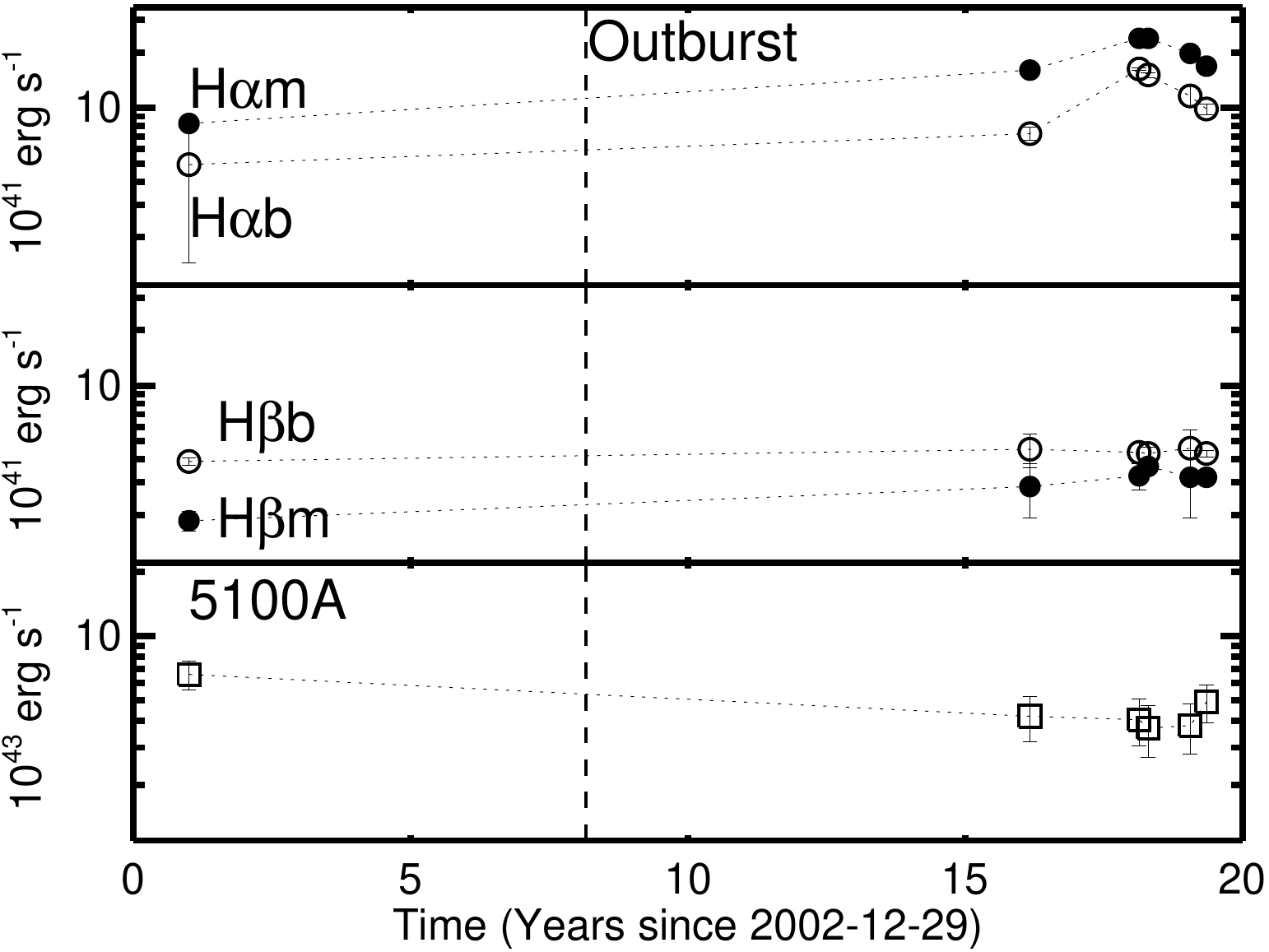}
 \caption{
  Variation of H${\alpha}$, H${\beta}$ and the continuum luminosity at 5100\AA~as a function of time. 
 Both broad and intermediate emission line components are shown. The dashed vertical line 
 indicates the time of peak luminosity of optical outburst observed on 2010 Feb 23 (D11). }
 \label{halc}%
 \end{figure}

 {In Figure \ref{haprofile}, we present the spectral decompositions of $\rm H\alpha$ line observed in six epochs. 
 The emission-line spectral fitting results are shown in Table 3.
 Interestingly, we found that the broad and intermediate component of $H_{\alpha}$ 
 emission line observed with P200 on 2018 are showing slight increase in flux with respect to the SDSS ones, 
 and the fluxes are rising to a peak during the P200/DBSP observation on 2020, by a factor of 
 $\sim$3.3 and $\sim$2.9 than that of SDSS ones (Table 3).
 Although the actual time for the onset of H$\alpha$ strengthening is not known, it remains strong at least 
 for $\sim$2 months based on the current data. 
 Our recent YFOSC spectrum taken on Jan 2021 and May 2021 shows that the broad H$\alpha$ flux 
 appears to have declined, though it is still a factor of $\sim$2 higher than that of pre-flare SDSS spectrum. 
 This strongly suggests that the H$\alpha$ brightening is a transient phenomenon. }
 We argue that such an H$\alpha$ flux variation is not due to observational effects, 
 as our simulations (see the details in Appendix A) have demonstrated that the changes in 
 the H$\alpha$ EW is at most at a level of 7\% under the conditions of five spectroscopic 
 observations (Table 3), i.e., for the seeing between 1.5\arcsec~and 3.3\arcsec, 
 and the slit width between 1.5\arcsec~and 2.5\arcsec.  
 Timely spectroscopy monitoring observations are highly encouraged to determine how long will the H$\alpha$ brightening last. 
  

 Since the optical broad emission lines (BELs) are believed to arise largely from the high-density gas 
 clouds in BLR (Broad Line Region) that is photoionized by an intense central continuum radiation, we would naively expect 
 correlated changes in the BEL strengths with the intrinsic variations in the incident ionizing 
 continuum flux. 
 This unique property between the continuum and emission line variations constitutes 
 the foundation of the reverberation mapping technique (RM, see Blandford \& McKee 1982), 
 and has proven observationally extremely useful to probe the spatial distribution and 
 kinematics of the BEL gas \citep[see][and references therein]{Peterson1993}. 
 \citet[][]{Greene2005} found that the luminosity of H$\alpha$ BEL is correlated with 
 the continuum luminosity at 5100\AA~for an ensemble of AGNs. 
 The best-fit relation suggests $L_{H\alpha}\propto L_{5100\AA}^{1.157}$, 
 with an rms scatter of $\sim$0.2 dex \citep[][]{Greene2005}. 
 If assuming \src follows this relation, the observed enhancement in the H$\alpha$ flux (combined broad and 
 intermediate components) on 2020 by a factor of $\sim$1.8 in \src with respect to that of 2018 
 would indicate a corresponding increase in the 5100\AA~continuum flux by $\sim$0.6$\pm$0.5 mag. 
 The actual 5100\AA~luminosity appears, however, to remain little changes with time during our follow-up spectroscopy 
 observing campaign, as shown in Table 3 and Figure \ref{halc}. 
 Such a reverse evolution between the H$\alpha$ line and continuum variations is even 
 extreme if compared with the pre-flare SDSS spectrum.

\begin{figure}[htbp]
 \centering
 \includegraphics[scale=0.43]{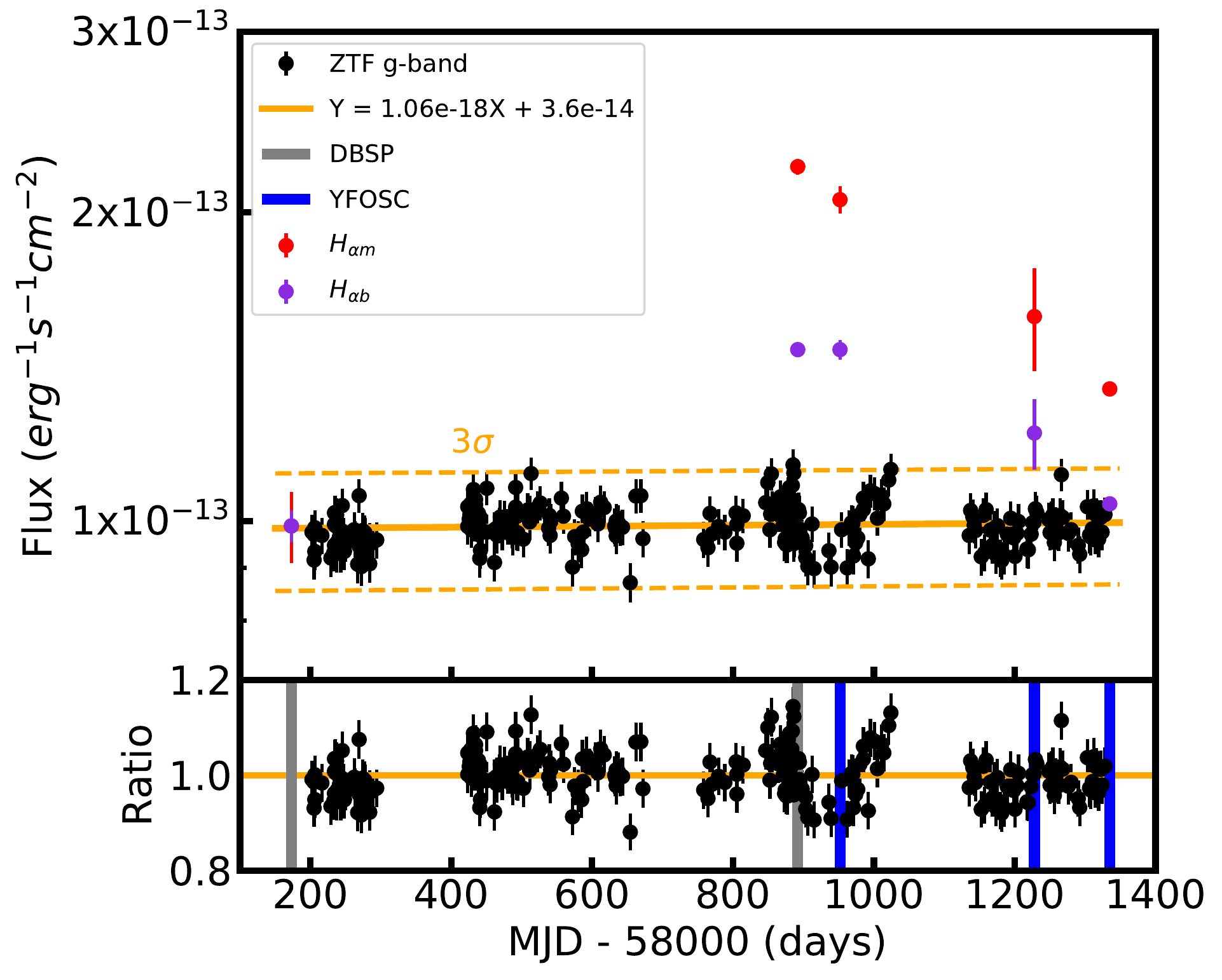}
 \caption{
  Comparison between the variation in H${\alpha}$ broad line (purple) and that of medium width component (red) 
 and the observed $g$-band continuum light curve (black points, {see also in Figure \ref{totallc})}. 
 The orange solid line represents the linear fit, suggesting that the $g$-band flux variability 
 can be described by a constant, with little evolution with time. 
 The dashed line indicates the 3$\sigma$ scatter of flux variability relative to its mean. 
 For clarity, the H${\alpha}$ flux has been normalized to the mean $g$-band flux on the date of Feb 24, 2018, 
 the start of our follow-up spectroscopic observations (thick vertical lines). 
 Bottom panel shows the ratios of the $g$-band fluxes to the best-fit mean. 
 }
 \label{ztfha}%
 \end{figure}

  Strictly speaking, the above comparison is unfair because the flux responsivity of H$\alpha$ BELs to the continuum 
 variations in individual AGNs might be different from the above luminosity-luminosity relation, such as AGNs with RM 
 observations for which the line responsivity typically has a scatter of less than 0.15 dex \citep[e.g.,][]{Shapovalova2012, Shapovalova2019}. To take into account 
 this effect, we compare in Figure \ref{ztfha} the variability of H$\alpha$ BELs with 
 {the continuum variability of \src at $g$-band observed by ZTF,  }
 which has the best sampling in light curve covering our follow-up spectroscopic observations (thick vertical lines). 
 For clarity, the H$\alpha$ light curves have been normalized to the mean value of $g$-band flux at the 
 date of our P18 observation. It can be seen that while the $g$-band flux remains constant with little evolution 
 with time, the variability amplitudes for H$\alpha$ BELs are much larger, above the 3$\sigma$ upper limit on the 
 $g$-band variability. Under the assumption of $F_{\rm line}\propto F_{\rm cont}^{\eta}$ in the RM model, the result 
 clearly suggests that the H$\alpha$ BEL variability in \src is not relevant to the continuum variation. 
 
 \subsection{Multi-wavelength photometric light curves}

\begin{figure}[ht]
 \centering
 \includegraphics[scale=0.4]{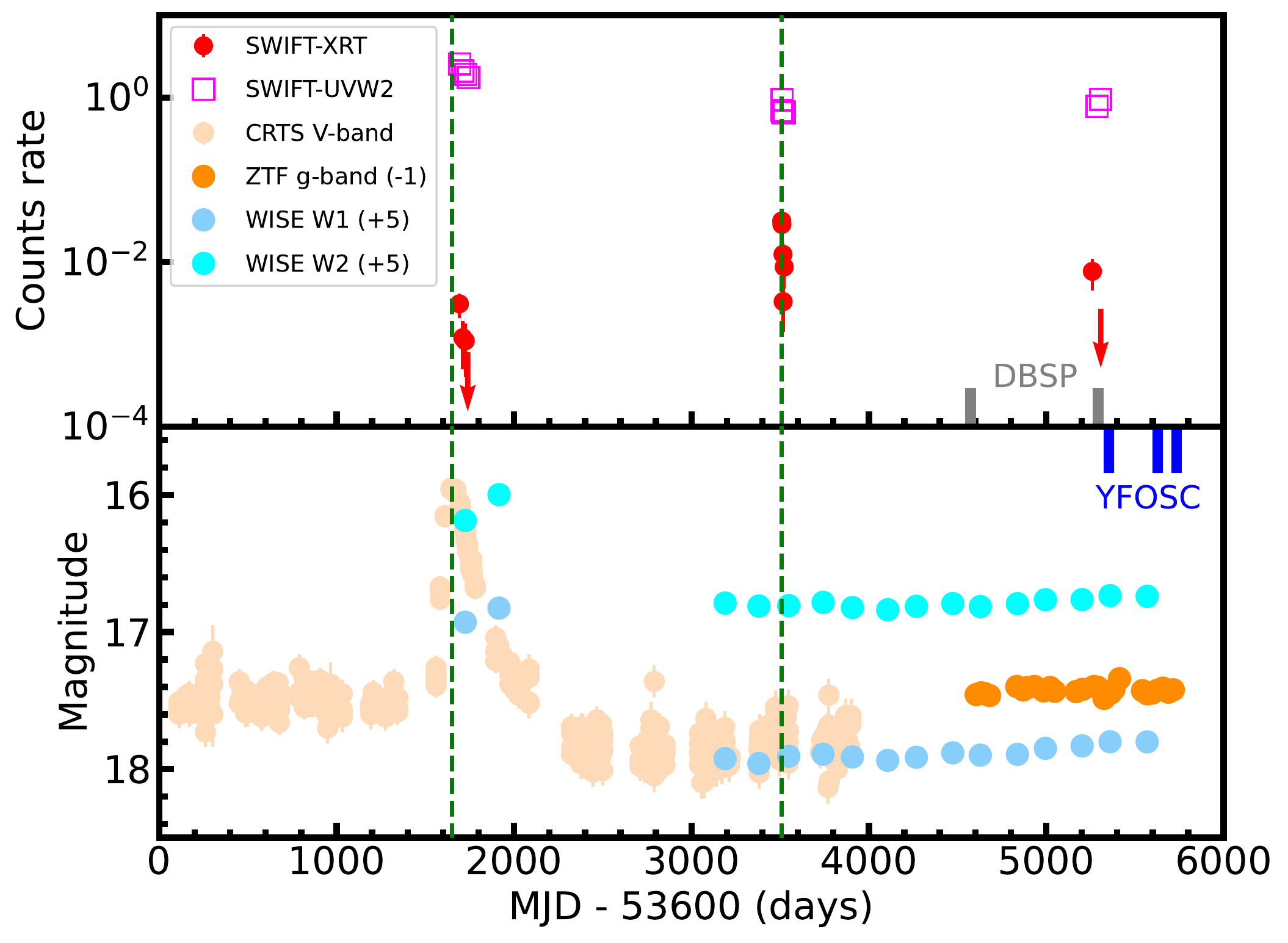}
 \caption{
  Multi-wavelength light curves showing the changes in continuum over the past 15 years for \srcs. 
 {\it Upper panel}: The UV (magenta) and X-ray (red) light curves observed from {\it Swift}. 
 {\it Bottom panel:} The light curve in the optical V-band from the CRTS DR2 (light orange), g-band from the ZTF DR2 (dark orange), and 
 in the MIR 3.4$\mu$m (cyan) and 4.6$\mu$m (light blue) from WISE. 
 The green dashed line represents the time of optical peak in 2010, and the X-ray peak in 2015, respectively. 
 The gray thick lines show the time of our DBSP spectroscopy observations, while the blue thick lines indicate 
 the time of YFOSC spectroscopy observations.  }
 \label{totallc}%
 \end{figure}

 \src was discovered by the CRTS survey on 2010 Feb (D11), but only optical photometry 
 up to $\sim$300 days since outburst is shown. 
 We retrieved the optical V-band photometric data from the CRTS website\footnote{http://nesssi.cacr.caltech.edu/DataRelease} 
 and updated to match the long-term light curve as presented in \citet{Graham2017}. 
 The V-band light curve is displayed in Figure \ref{totallc}. 
 It can be seen that the source clearly displays two distinct states in the optical band: 
 a long quiescent state and an outburst state which differs in brightness by $\sim$1.5--1.8 mag. 
 The quiescent brightness decreases by $\sim$0.3 mag after the outburst, suggesting a slight 
 change in the accretion rate. 
 In addition, we built MIR light curves by collecting photometric data at 3.4$\mu$m (W1) and 4.6$\mu$m (W2) from the 
 WISE (Wide-field Infrared Survey Explorer) survey up to  Dec 2020.  
 Details on the WISE photometry and lightcurve construction are given in \citet{Jiang2016, Jiang2019}.  
 The MIR flares are clearly presented soon after the optical outburst (MJD 55325), probably 
 due to the dust reprocessed emission from the circumnuclear region. 
 The MIR flare is followed by a fading back to a long quiescent level, similar to that observed in the optical. 
 The trend is confirmed with the recent g-band light curve from the ZTF (Zwicky Transient Facility) survey\footnote{https://irsa.ipac.caltech.edu/applications/ztf/}, 
 which displays a plateau over a period of $\sim$3 years (up to May 2021). 
  We also show the long-term X-ray and UV light curves observed by {\it Swift}, though the 
   sampling of the data is much poor. 

 \subsubsection{Analysis of UV to optical SED during outburst}
  We fitted a blackbody model ($B_{\nu}=\frac{2h\nu^3}{c^2}\frac{1}{e^{h\nu/kT}-1}$) to the UV-to-optical photometric data from the 
 \swift UVOT observations during the period of optical burst, to put constrains on the luminosity, temperature 
 and radius evolution of UV and optical emission of \srcs. 
 \swift UVOT observations cover a wavelength range from $\sim$1900\AA~to 5500\AA~, which is sensitive to probe the UV-to-optical SED of 
 typical optical TDEs \citep[e.g.,][]{vanVelzen2020}. 
 In addition, the simultaneous UVOT observations at six filters are helpful to migrate the uncertainties on 
 SED fittings due to the variability in individual bands. 
 We utilized the archival as well as our own target-of-opportunity \swift observations performed on Jan 2020 and Feb 2020 to 
 estimate the quiescent host emission. The host emission at each filter is estimated as the mean of flux between the two epochs, 
 which was then subtracted from the total UVOT measurements to obtain the transient photometry during the period of the outburst 
 (and the errors on host flux were propagated). The transient photometry was then corrected for the Galactic extinction of $E(B-V)=0.015$ mag \citep{Drake2011}. 
 We found that the host-subtracted, Galactic extinction-corrected UV-to-optical SED of \src can be well 
 described by the blackbody model, with a reduced $\chi^2$/dof of $\sim$0.8-0.9 {(see Appendix B for the fitting results)}.

 \begin{figure}[htbp]
 \centering
 \includegraphics[scale=0.55]{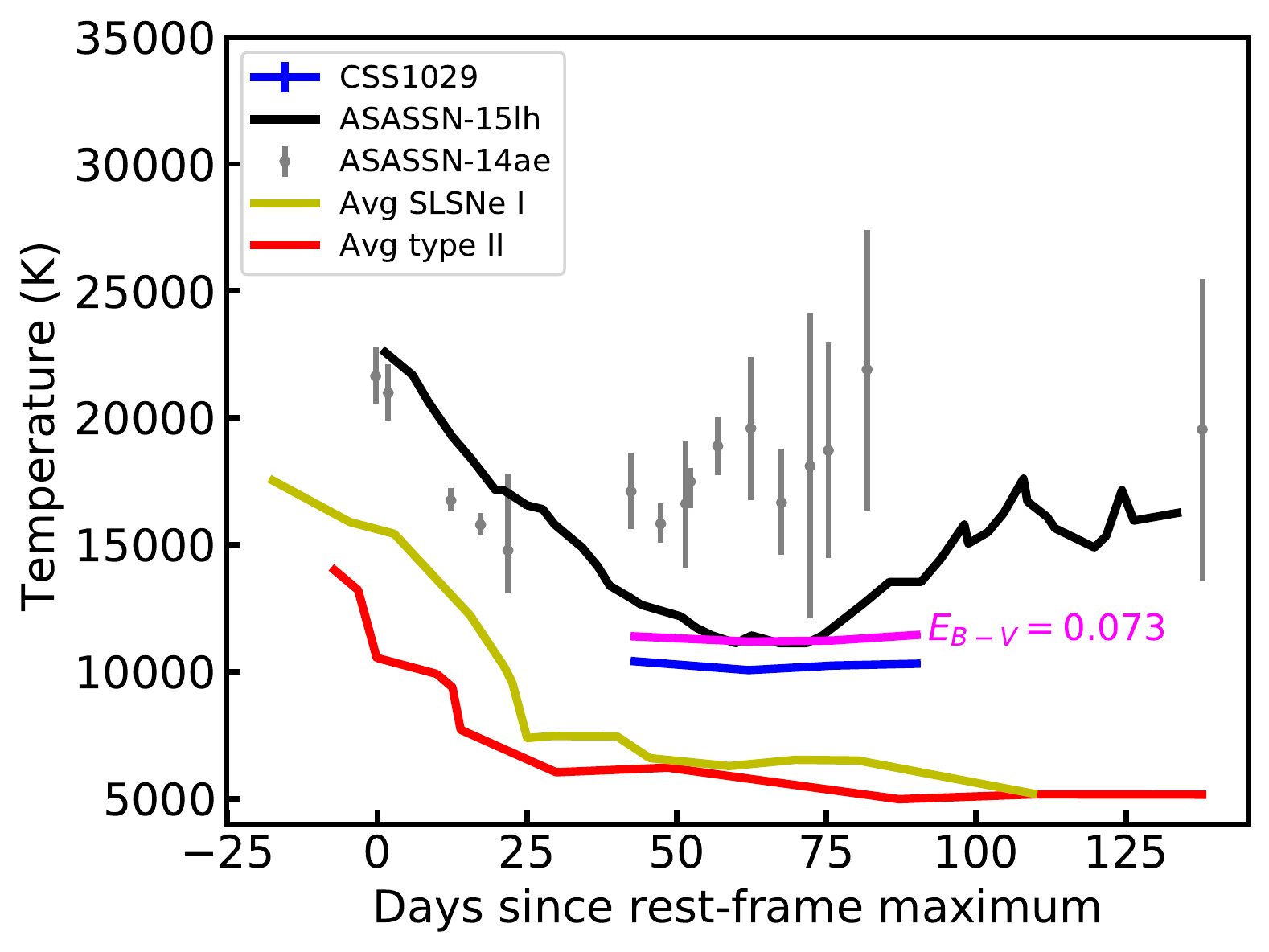}
 \caption{
 {\it Upper panel}: Temperature evolution of UV-optical emission of \srcs, as compared with that of superluminous supernovae (SLSNe) and type II SNe \citep{Inserra2018}. 
 The blue curve shows the blackbody temperature measured from the photometric data corrected for the Galactic dust extinction ($E_{B-V}$=0.015), while 
 the purple curve represent that corrected for both the Galactic and internal dust extinction ($E_{B-V}$=0.073). 
 The latter is estimated from the Balmer decrement of narrow emission lines \citep{Drake2011}. 
 {The temperature evolution of \src appears similar to that of optical TDE ASASSN-14ae \citep[gray points,][]{Holoien2014}, 
 as well as the peculiar SLSN ASASSN-15lh \citep[black curve,][]{Dong2016}. The latter has been suggested to be 
  more consistent with a TDE based on recent observations \citep{Leloudas2016}.}
  }
 
 \label{temp}%
 \end{figure}

  The SED fitting results indicate a flat evolution of blackbody temperature between $\sim$40-90 days since the optical peak, 
 with a mean temperature of $\sim$$1.03\times10^4$ K, as shown in the blue curve in Figure \ref{temp} (upper panel). 
 We note that the temperature fit to UV-to-optical SED is very sensitive to internal extinction. 
 Although \citet{Drake2011} suggest that the host galaxy extinction of \src is a small effect, we estimated 
 a potential internal extinction with the narrow line Balmer decrement as measured from the preflare 
 SDSS spectrum \citep{Drake2011}. 
 By summing the Galactic extinction, we found a total extinction $E_{B-V}$=0.073 assuming an ``average"
  reddening for the diffuse interstellar medium of R(V)=3.1. 
  Correcting for the total extinction to the UV-optical SED observed by {\it Swift}, the best-fit temperature 
  increases to $\sim$$1.13\times10^4$ K (purple curve in Figure \ref{temp}, upper panel), 
 well in agreement with the blackbody component seen in the spectra \citep{Drake2011}. 
 By comparing with the average trend of temperature evolution of the hydrogen-poor  
 superluminous supernovae (SLSNe I) and type II SNe \citep[e.g.,][]{Inserra2018}, we found that \src displays 
 considerably higher temperature (by a factor of $\sim$1.5) than the SNe at the same epoch 
 ($\sim$40-90 days since the peak). The dramatic 
 temperature evolution for the SNe is likely resulting from cooling due to rapid expansion 
 and radiative losses. 

 \subsubsection{{\tt MOSFIT} fittings to the optical light curve of outburst}
 
\begin{figure}[]
 \centering
 \includegraphics[scale=0.49]{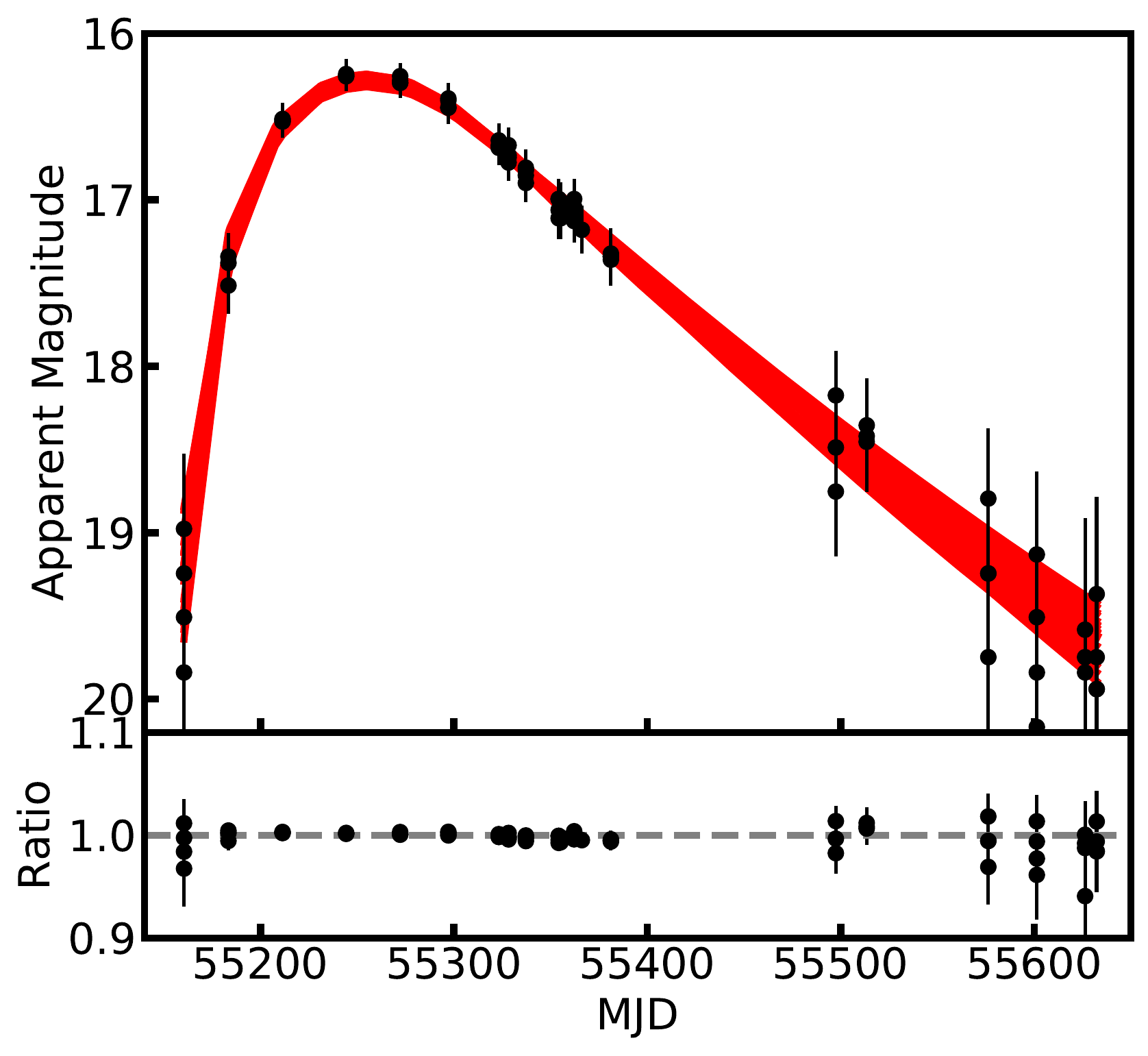}
 \caption{
  Host-subtracted V-band light curve of \src (filled circles), and the best model realizations from {\tt MOSFIT} (red curves).
 The most relevant best-fit model parameters are listed in Table 4.  }
 \label{mosfit}%
 \end{figure}

  We further explored whether the optical light curve of outburst can be fitted with the Monte Carlo 
 software {\tt MOSFIT}, which was recently applied to model the light curves of TDEs \citep{Mockler2019}. 
 This has not been done in the previous work (D11). 
 The TDE model in {\tt MOSFIT} assumes that emission produced within an 
 elliptical accretion disk of a TDE is partly reprocessed into the UV/optical by an optically thick layer 
 \citep{Guillochon2018}. 
 We run {\tt MOSFIT} using a variant of the {\tt emcee} ensemble-based Markov Chain Monte Carlo routine, 
 until the fit has converged by reaching a potential scale reduction factor of $<$1.2 \citep{Mockler2019}. 
 In Figure \ref{mosfit}, we show the host-subtracted V-band light curve, and 
 an ensemble of model realizations from {\tt MOSFIT}. The model is able to reproduce the data quite well, 
 including the stages of the rise to peak, near the peak and steady decline at later times. 
 To quantify how well the various combinations of parameters in the modelling, {\tt MOSFIT} 
 uses the Watanabe-Akaike information criteria \citep[WAIC;][]{Gelman2014}\footnote{WAIC is a widely applicable Bayesian criteria, and evaluated as $\rm WAIC=\langle log$$p_{n}\rangle$-var(log$p_n$), 
 where $\langle logp_{n}\rangle$ is the mean of log likelihood score and var(log$p_n$) is its variance.}. 
 For \srcs, we found $\rm WAIC=119$, which is fairly consistent with the best-fit values for a sample of 
 optical TDEs \citep{Mockler2019}, suggesting that the fitting result is acceptable. 

\begin{figure}[htbp]
 \centering
 \includegraphics[scale=0.29]{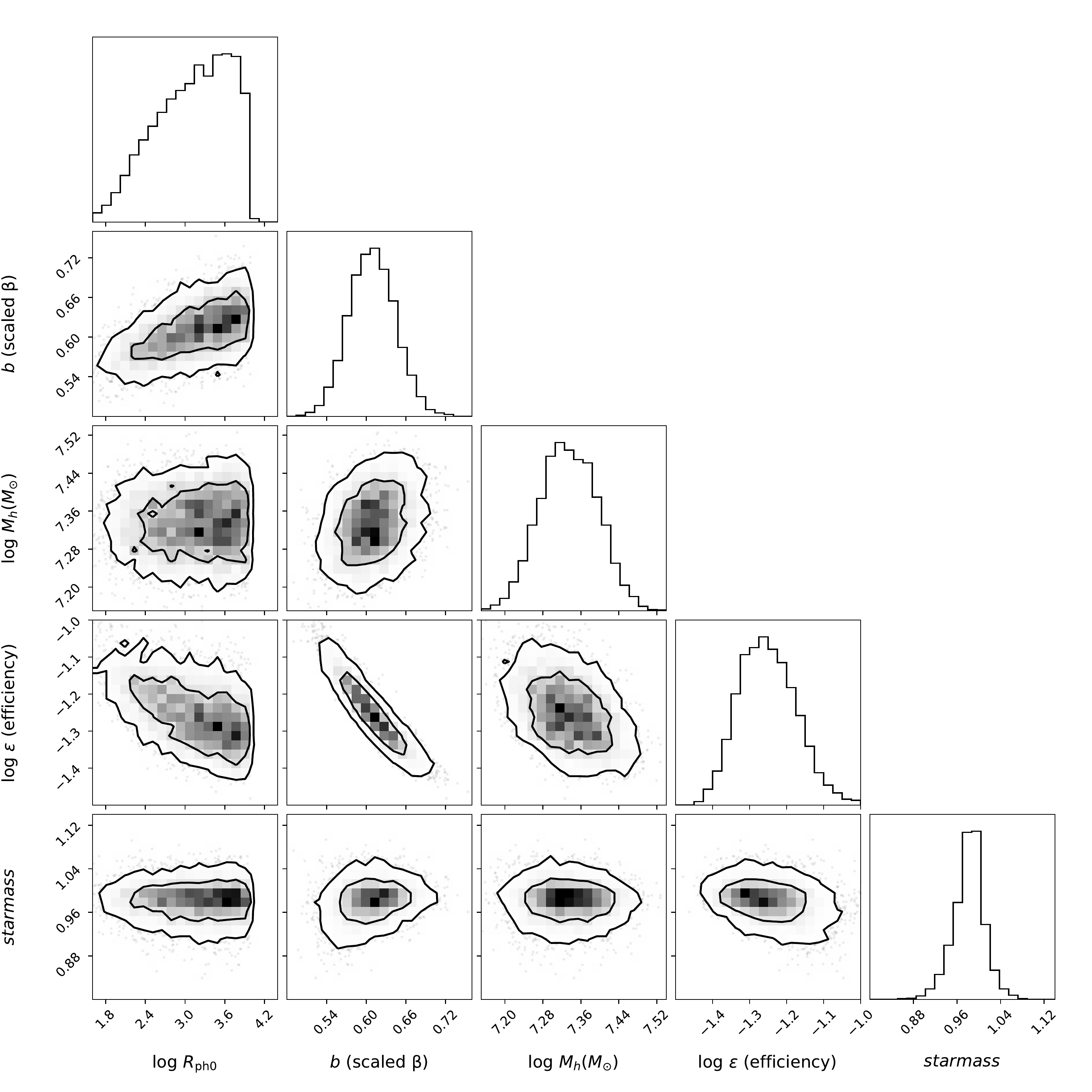}
 \caption{
  Posterior distributions of model parameters from the fit 
 to the V-band light curve of \srcs, shown in Figure \ref{mosfit}. 
 The contours (from inside to out) indicate the 1$\sigma$ and 2$\sigma$ confidence intervals 
 for two parameters of interest, respectively.  
 }
 \label{para}%
 \end{figure}

 The best-fit parameters of the TDE model with the corresponding systematic and statistical errors at 1$\sigma$ 
 confidence are shown in Table 4. Figure \ref{para} shows the posterior probability distribution of the most relevant 
 model parameters in the fit. The best-fit model is that of a black hole of $2.1\times10^{7}$\msun~disrupting a star of 
 0.98 \msun. 
 This black hole mass is consistent within errors with the mass estimated by \citet{Blanchard2017} using the 
 $H_{\alpha}$ broad emission line. 
 In comparison with other TDEs modeled in \citet[][]{Mockler2019}, the stellar mass for \src is larger than most other TDEs 
 where a disrupted star with mass near 0.1\msun~is preferred. 
 As Table 4 shows, the best-fit model from {\tt MOSFIT} indicates that the star was likely partially 
 disrupted as the impact parameter $b<1$ \citep[][]{Mockler2019}. 
 This is similar to the case in the TDE AT2018hyz \citep[][]{Gomez2020}, 
 which has a $b=0.4$.



 \begin{table}
  \centering
   \caption{Best-fit model parameters from {\tt MOSFIT}. }
    \renewcommand\arraystretch{1.3}
   \setlength{\tabcolsep}{5mm}{
  \begin{tabular}{c c c c}
    \hline 
    \hline 
Parameter & Prior & Value & Units \\
    \hline  
      $\rm log (R_{ph0})$ &  [-4, 4] & $3.18^{+0.53}_{-0.69}$ &  \\
      $l$ & [0, 4] & $1.52^{+0.43}_{-0.39}$ &  \\
       $\rm log (T_{viscous})$ & [-3, 5] &$-0.67^{+1.3}_{-1.1}$ & days \\    
       b (scaled $\beta$) &  [0, 2] & $0.61^{+0.04}_{-0.03}$ &  \\
       $\rm log (M_{BH})$ & [5, 8] & $7.33^{+0.06}_{-0.06}$ &  $M_{\odot}$ \\
        $\rm log \varepsilon$ & [-2, -0.4] & $-1.26^{+0.07}_{-0.07}$ & \\
        $starmass$  & [0.01, 10] & $0.98^{+0.02}_{-0.03}$ & $M_{\odot}$ \\
         $\rm log \sigma$ &  [-4, 2] & $-1.08^{+0.11}_{-0.11}$& \\    
\hline
\hline
  
\end{tabular}}
 \tablefoot{The parameter $R_{ph0}$ and $l$ is the radius normalization and power-law exponent 
     to compute the luminosity-dependent radius of the photosphere in {\tt MOSFIT}; 
      $T_{viscous}$ is the viscous time of the gas accretion onto the black hole; 
      $b$ is a proxy for $\beta$, $\beta = R_{t}/R_{p}$, where $R_{t}$ refers to the tidal disruption radius 
     and $R_{p}$ refers to the pericenter radius. Full stellar disruption corresponds to $b=1$; 
     $M_{h}(M_{\odot})$ is the black hole mass; $\varepsilon$ is the efficiency parameter 
     that converts the input fallback rate of material to luminosity; 
     starmass is the mass of disrupted star; 
     $\sigma^2$ represents the additional variance 
     in the Bayesian analysis to make $\chi^2_{r}$=1. The errors are only for the fitting statistical uncertainties, 
     and do not include the systematic uncertainties from {\tt MOSFIT} \citep{Mockler2019}. }
  
\end{table}

 \section{Discussion and Conclusion}

 \subsection{Nature of optical outburst}

 Since its discovery, the nature of optical outburst shown in the CRTS V-band lightcurve (Figure \ref{totallc}) 
 is in dispute. D11 explored an origin of optical outburst as an AGN flare, a TDE, or a Type IIn SLSN, and argued 
 for the last option that has to occur within 150 pc of the nucleus of the galaxy. 
 {The normal AGN variability was ruled out as such a rapid (duration of $\sim$1 yr), energetic optical flare 
 (total integrated energy of $\sim10^{52}$ erg) is rarely seen in the light curves of NLS1 \citep[D11;][]{Kankare2017}. 

 On the other hand, the MIR outbursts in \src are clearly detected with WISE. 
 In both the cases of SNe and TDE, strong MIR flares are generally thought to be thermal emission from the dust heated 
 by the nuclear optical outburst.
 The optical emission from the outburst can sublimate the dust within a certain range and form a dust cavity, 
 the radius of which can be described by the sublimation radius $r_{\rm sub}$.
 Assuming a dust sublimation temperature of 1800 K and a grain radius of 0.1 $\mu$m, the peak bolometric luminosity of 
 the outburst of $4\times10^{44}$ erg s$^{-1}$ in CSS1029 suggests a $r_{\rm sub}$ of $\sim$0.3 light-years\footnote{
 Here we used the formula given in \citet{Namekata2016} to estimate the sublimation radius 
 \begin{equation}
 r_{\rm sub}=0.121pc\left(\frac{L_{\rm bol}}{10^{45} erg s^{-1}}\right)^{0.5}\left(\frac{T_{\rm sub}}{1800 K}\right)^{-2.804}\left(\frac{a}{0.1\mu m}\right)^{-0.51}.
 \end{equation}}.
 The IR emitting dust has a distance $r\gtrsim r_{\rm sub}$ from the central source, causing the IR emission delayed 
 relative to optical emission.
 If the dust is heated directly by central UV/optical radiation, the time delay $\tau\sim r/c$ would be expected as several months.
 Conversely, if the dust is shock-heated by the material moving outward at a speed of $v_{\rm out}$ (typically $\sim$0.02-0.1 c), 
 from either the ejecta for SNe or outflow for TDE, 
 the time delay $\tau\sim r/v_{\rm out}$ would be several years or longer.
 Since the MIR outburst in \src was detected several months after the optical peak, the dust should be heated directly by radiation. 
 The possibility of non-thermal radiation from a relativistic jet to cause the MIR flare can be ruled out, because the galaxy 
 is radio quiet ($f_{\rm 6cm}/f_{4400\AA}\simlt1$, D11). 
 The host-subtracted peak MIR luminosity of CSS1029 is as high as of $1.26\times10^{44}$ erg s$^{-1}$.
 As shown in Figure \ref{lumevo}, the peak MIR luminosity is among brightest ones for MIR nuclear transients, 
 which is about two orders of magnitude higher than typical SNe explosions \citep{Jiang2019, Jiang2021, Szalai2019}, 
 and one order of magnitude higher than the rare population of SLSNe explosions, such as ASASSN-15lh \citep{Dong2016}.
 The peak MIR luminosity is close to those of several TDE candidates, including F01004-2237 \citep{Dou2017}, 
 Arp 299B-AT1 \citep{Mattila2018} and PS1-10adi \citep{Jiang2019}.
 Thus the high MIR luminosity prefers to the TDE interpretation of the outburst.
 Note that because of the sparse sampling of the MIR photometry during the outburst phase, the 
 true peak luminosity for \src may be even higher.
 Furthermore, the optical peak luminosity matches the Eddington luminosity of the SMBH \citep{Blanchard2017}, supporting 
 its association with accretion.

 \begin{figure}[htbp]
 \centering
 \includegraphics[scale=0.4]{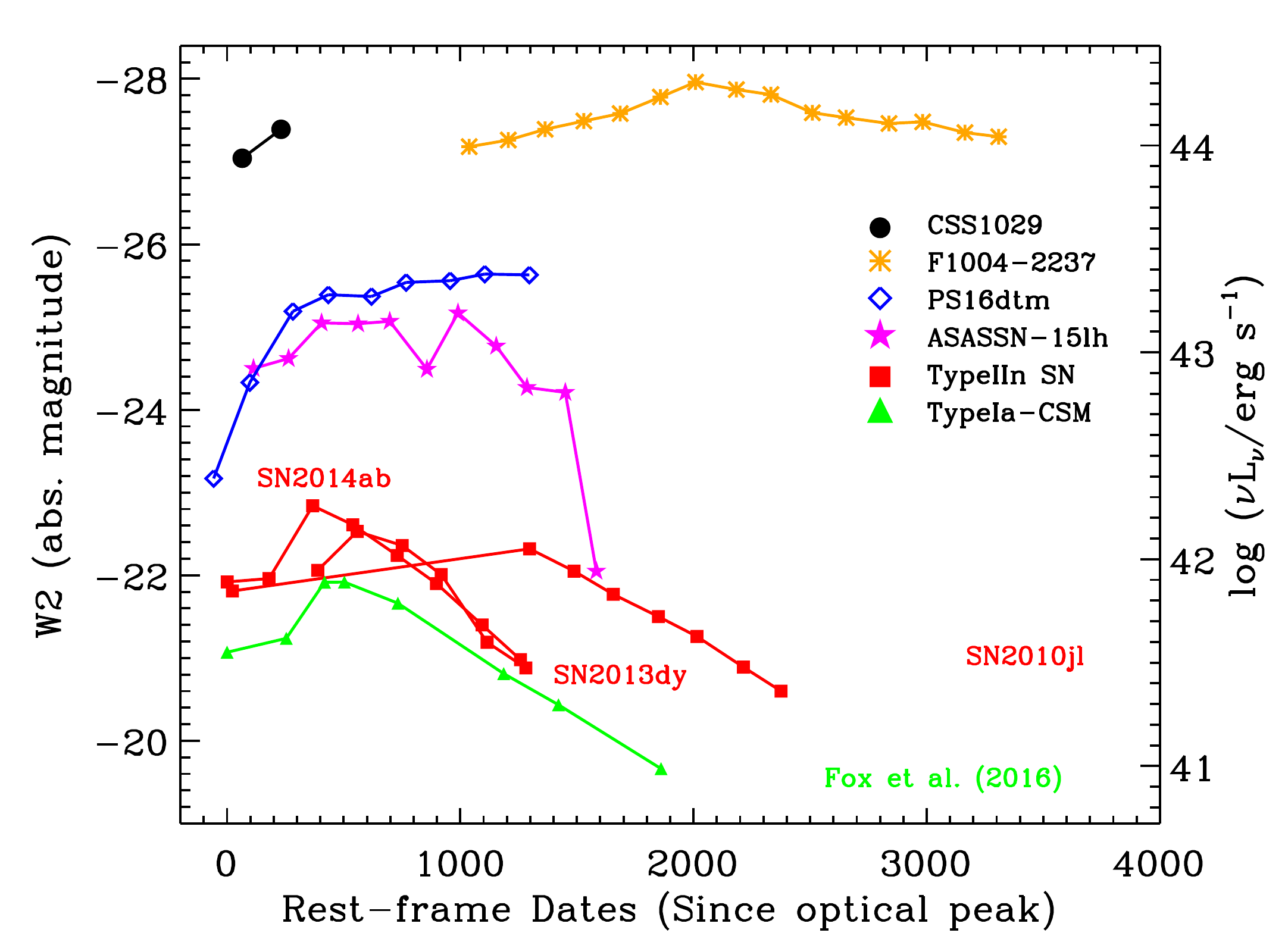}
 \caption{
 Host-subtracted WISE W2 (4.6$\mu$m) magnitude for \src during outburst. For comparison, we also plot the evolution of peak 
 MIR luminosity of the TDE F01004-2237 \citep{Dou2017} and PS16dtm \citep{Blanchard2017}, SLSNe \citep{Jiang2019} and the supernovae with 
 an excess of MIR emission, i.e., Type Iax SN 2014dt \citep{Fox2016}.
  }
 
 \label{lumevo}%
 \end{figure}

 In Section 3.5.1, we have shown that \src exhibits a roughly constant temperature of $T\simeq11300$ K 
 between $\sim40-90$ days since the optical peak. 
 This is neither expected from the photosphere evolution of type II SNe, nor observed for hydrogen-poor SLSN \citep{Inserra2018}. 
 For a typical SNe both its temperature and bolometric luminosity drop roughly as power-laws with time during the early 
 stage ($\sim1-3$ weeks), possibly as a result of cooling due to rapid expansion and radiative losses 
 \citep[e.g.,][]{Faran2018}. Once the hydrogen recombination becomes important, which dominates the drop 
 in the optical depth of the photosphere expansion, the temperature remains almost constant at $\sim$6000-7000 K. 
 With a temperature that is higher than 
 averaged SNe by a factor of $\sim1.5$ at the same epoch (Figure \ref{temp}), this suggests that \src might be different from 
 typical SNe explosions.
 However, the TDE candidate ASASSN-14ae showed a very similar temperature evolution to \srcs. 
 In fact, most TDEs discovered in the optical bands exhibit relatively flat temperature evolution 
 over several months since peak \citep{vanVelzen2020, Holoien2021}. 
 The constant temperature evolution could be explained if the UV/optical emission is powered by 
 the reprocessed accretion radiation by unbound TDE outflow \citep{Metzger2016}.  
 This similarity between the evolution of \src and TDEs suggests that they might be due to the same mechanism. 
 
 On the other hand, the temperature for \src is on the low end of the temperature
 range observed for TDEs (of $\sim1-4\times10^4$ K). \citet{vanVelzen2020} found a 
 significant anti-correlation between the blackbody radius and temperature, following the 
 relation $L_{\rm bb}\propto R^2T^4$. By fitting to the UV to optical SED, 
 we found that the blackbody radius of \src is $\sim6-8\times10^{15}$ cm (Figure \ref{bbodyevo} in Appendix B), 
 on the high end of the radius range observed for TDEs \citep{Holoien2021}. If assuming that the blackbody
photospheric radius is proportional to the size of accretion disk in the context of TDEs, 
the large blackbody radius of \src could be attributed to the higher mass of disrupted star \citep{vanVelzen2020}. 
This is consistent with the results of {\tt MOSFIT} fittings to the V-band light curve (Table 4), 
where a black hole of $2\times10^{7}$\msun~disrupting a star of 0.98 \msun~is inferred. 


 Note that D11 argued against a TDE explanation primarily because the blackbody temperature and flux decline rate were 
 inconsistent with theoretical predictions from early studies of TDEs. However, as we shown in Section 3.1, the 
 exponential decline rate that can best describe the outburst evolution of \src has also been observed in many TDEs 
 that are discovered in modern optical surveys \citep[e.g., ][]{Holoien2014, Holoien2016, Shu2020}. 
 In addition, as shown in Section 3.4, our spectroscopy follow-up observations have revealed the brightening 
 of the broad $\rm H\alpha$ flux, about a decade after the optical outburst. 
 Although a broad $\rm H\alpha$ emission component has been observed to emerge in the late-time spectra 
 of some SLSNes \citep{Gal2019}, the time-scale for the appearance is much shorter, typically $\sim$1 yr. 
 This makes the origin of late-time $\rm H\alpha$ brightening in \src distinct from what is expected in SLSNe, 
 likely related to a TDE.   
Given the high peak MIR luminosity and relatively constant temperature evolution of $T\sim11000$ K 
over $\sim$40-90 days since optical peak, we conclude that a TDE origin is preferred for the optical 
outburst for \src and the scenario with a superluminous SNe explosion seems disfavored. 

}


 \subsection{Possible origins for the X-ray brightening}







 By analyzing the follow-up observations with \swift taken on 2015 and 2020, 
 we found an X-ray brightening of $\simgt$30 in \src relative to its previous low state in 2010, 
 followed by a flux fading within about two weeks. 
 The appearance of bright X-ray emission at later times and its rapid decline in flux within a short period 
 appear to be an unusual property of \srcs. 
 Since it was established previously that the optical spectrum of \src is that of NSy1 (D11), this 
 may suggest that the X-ray brightening could be attributed to the AGN activity. 
   This is supported by the detection of a hard X-ray component with 
 2-10 keV luminosity of $8.9^{+6.7}_{-5.1}\times10^{42}$\erg (Section 3.2), and 
 a best-fit photon index of $\Gamma\sim2.1$ typically for NSy1s. 
  However, we note that the large-amplitude X-ray variability in \src is probably distinct from 
 the rapid AGN variability seen on shorter time-scales of 
 {hours to days \citep[e.g.,][]{Uttley1999}\footnote{
 From the extinction-corrected {\sc [O iii]}$\lambda$5007 luminosity of $2.4\times10^{41}$\erg from the SDSS spectrum (D11), 
 we roughly estimated the ``time-averaged" 2-10 keV luminosity of $3.0^{+0.8}_{-0.6}\times10^{42}$\erg \citep[e.g.,][]{Lamastra2009}. 
  This is a factor of $\sim$3 higher than the upper limit of hard X-ray luminosity obtained in 2010,  
 suggesting that \src could be in a historically low X-ray state during the optical outburst. }.  
}

 As mentioned in Section 3.3, the UV emission shows little variability during the X-ray flaring, 
 with maximum variability amplitude of $\sim$50\%, which is mild compared with the X-ray one. 
 Moreover, as shown in Figure \ref{totallc}, we do not find any coincident variability in either optical or MIR with the X-ray 
 flaring in 2015. Note that similar X-ray flares have been observed in the AGN IC 3599 which could be attributed to an 
 instability in a radiation-pressure dominated accretion disk \citep{Grupe2015}, 
 or multiple tidal disruption flares with the recurrence time of 9.5 yr \citep{Campana2015}. 
 However, in this case a large optical outburst was also observed prior to the X-ray one, 
 making it physically different from what is observed in \srcs.  
 This suggests that the scenario that the dramatic change in AGN accretion rate causing the extreme 
 X-ray variability in \src seems unlikely. 
 The microlensing by stars in foreground galaxies seem also unable to explain the observed X-ray behavior, 
 as the model predicts a variability that is independent of wavelengths \citep[e.g.,][]{Lawrence2016}.

 The apparent lack of coincident optical/UV variability with X-ray flaring in \src is somewhat similar to that found in 
 some NSy1s with extreme X-ray variations \citep[e.g.,][]{Bachev2009, Miniutti2012, Grupe2008, Buisson2018, Grupe2019}. 
  It is suggested that the absorber with a non-unity covering fraction at large scale can cause the X-ray variability, 
  but it requires to be dust-free so as not affecting the UV or optical emission. 
 Assuming a single large cloud passing over the X-ray source, we can use the observed duration of 
 X-ray flux declining in 2015 (at least 10 d, Table 1) to constrain the location of the absorber. 
 The crossing time for an intervening cloud on a Keplerian orbit can be estimated as  
 \begin{equation}
 t_{\rm cross}=(GM_{\rm BH})^{-1/2}r_{c}^{1/2}D_{\rm src}	
 \end{equation}
 where $r_{c}$ is the distance of the cloud to X-ray source, $D_{\rm src}$ is size for the X-ray source 
 and $t_{\rm cross}\approx D_{\rm src}/V_{K}$ \citep[$V_{K}$ is Keplerian velocity,][]{Risaliti2007}.   
 For a BH mass of $7.9\times10^6$\msun~\citep{Blanchard2017} and a typical diameter of 10$R_{\rm G}$ for the X-ray 
 source \citep[e.g.,][]{Fabian2009}, we find the location of intervening cloud $r_{c}\simgt5\times10^{18}$ cm, 
 in order to have an obscuration that lasts longer than 10 d.  
 This corresponds to $\sim$1.6 pc, which is suggestive of a cloud that lies outside of the BLR. 
 Note that this distance is much larger than the typical dust sublimation radius in AGNs \citep[$\sim0.1$pc,][]{Jiang2019}, 
 making the explanation for the presence of dust-free clouds at such a large distance challenging. 
  Therefore, the scenario that X-ray brightening in \src due to the movement of dust-free absorbing 
 gas at large scale seems disfavored. 

 
 {Having established that TDE is plausible to power the optical outburst, 
 we discuss the possibility of whether a TDE could contribute to the X-ray variability in \srcs. 
Hydrodynamics simulations have shown that the debris stream of a TDE can affect the accretion properties 
of pre-existing AGN \citep{Chan2019}. 
\citet{Blanchard2017} presented the observation of PS16dtm, a TDE occurred at the nucleus of a NSy1, which 
shares many similarities with \srcs. 
\citet{Blanchard2017} suggested that the TDE can lead to the dimming of the X-ray emission from the pre-existing AGN, 
possibly due to the obscuration of X-ray emitting region by stellar debris. 
They further predicted that the source will become X-ray bright again approximately 
a decade since the TDE, though it has not been confirmed yet. In comparison with PS16dtm, the relatively 
faint X-ray flux of \src observed in 2010 could simply due to the partial obscuration of X-ray emitting region by 
accreting stellar debris, and the brightening in 2015 could be attributed to the decrease in the optical 
depth of absorbing gas as the accretion rate declines. 

On the other hand, \citet{Ricci2020} found the clear evidence that the hard power-law component (produced in X-ray corona) 
disappeared after the UV/optical outbursts in the AGN 1ES1927+65, i.e., drops in luminosity by more than three orders of 
magnitude. Interestingly, such a power-law component reappeared about one year later. \citet{Ricci2020} invoked the 
TDE scenario to explain the drastic X-ray variability, which can deplete the innermost regions of accretion flow \citep{Chan2019}, 
hence disrupting the magnetic field powering the X-ray corona. 
If the relative X-ray faintness of \src in 2010 is due to the destruction of AGN X-ray corona, 
its X-ray luminosity ($L_{\rm 0.3-2 keV}\sim3.7\times10^{42}$ \erg) appears too high in comparison to what is observed in 1ES1927+65. 
In this case, the observed X-ray emission of \src in 2010 could be attributed solely to the TDE.

 Alternatively, 
 several works have invoked a partial covering model to account for extreme X-ray variability behaviors 
 of AGNs, especially those high accreting systems \citep{Luo2015, Liu2019, Ni2020}.
 In this model, the central X-ray source could be partially obscured by a thick inner disk or outflow if 
 it intercepts the line of sight. 
 Similar scenario has been proposed to explain the faintness of the X-ray emission in optical TDEs by the orientation 
 effect where the X-rays from the inner disk is obscured by dense outflow gas \citep{Dai2018}.  
 The slight changes in the covering factor with respect to the X-ray source would result in an 
 X-ray variability, but the bulk UV/optical continuum at outer disk remains little affected. 
 Indeed, \citet{Blanchard2017} estimated that the Eddington ratio is close to 1 for \src when it reaches 
 to the peak luminosity of outburst on 2010, indicating the central BH is accreting at a high rate during this epoch. 
 Simulations have suggested that geometrically thick inner accretion disks and associated dense outflows will be formed when 
 accretion rate is high enough \citep{Jiang2014, Jiang2019}.   
 This is suggestive of the possibility that a thick disk in the inner region formed in \src 
 plays a role in partially obscuring central X-ray source, explaining the low X-ray flux observed in 2010. 
 The X-ray brightening observed at later times may be due to a decrease in the thickness or covering factor of 
 the inner disk with accretion rate (radiation luminosity), possibly induced by the TDE process. 
 As shown in Figure \ref{totallc}, the baseline AGN optical emission decreased by a factor of $\sim$0.3 mag after the outburst 
 seems to be in accordance with this scenario.
}

 \subsection{H$\alpha$ anomaly and implications for the BLR evolution}

{In Section 3.4, our analysis suggests transient H$\alpha$ brightening in \src that might not be relevant to the continuum variation.}
 Although we cannot locate spectra taken between 2018 and 2020, the most recent photometric observations 
 by \swift show that the amplitude of continuum variability at UV/optical bands was not as large 
 as that expected from the H$\alpha$ brightening (Table 1). 
 The lack of corresponding ionizing continuum variability is further supported by  
 the WISE light curves at MIR bands (Figure \ref{totallc}), 
 which suggest no recent outburst even for the largely unobservable extreme-UV variation up to Dec 2020. 
 Otherwise an MIR echo signal might be detected. 
 We argue for that the "non-responsivity" of ionizing continuum variability is not due to the 
 contamination of constant host emission, as the optical continuum is dominated by the AGN 
 emission in the spectroscopic data (Section 3.4). 
{The clear flux decline in the V-band light curve by $\sim$0.3 mag (Figure \ref{totallc}) is also suggestive of 
 the dominant AGN contribution to the optical photometry.} 
 Therefore, the observed H$\alpha$ brightening is an anomalous behavior of the BEL variations in the context of RM model. 
 Note that transient anomalous phenomenon has been reported in other AGNs from the RM campaign \citep[][]{Goad2016, Gaskell2021}, but 
 most are characterized by a significant deficit in the flux of BELs when continuum is brightened,  
 which is clearly different from the H$\alpha$ anomaly observed in \srcs.

 {As no clear evidence for a temporary increase in the ionizing continuum incident upon BLR clouds is found 
 during/before the anomaly, it could be possible to explain the enhanced H$\alpha$ emission by an increase of 
 gas density or excitation within BLR itself.  }
 Although the formation processes of the BLR are still unknown, 
 \citet[][]{Czerny2011} suggested that the BLR is a failed dusty wind from outer accretion disk. 
 \citet[][]{Baskin2018} refined this model by exploring the dust properties as well as implied BLR structure 
 in more detail, 
 and proposed a dust inflated accretion disk as the origin of the BLR. 
 The dusty disk wind scenario may not be able to explain the H$\alpha$ anomaly, as 
 it requires a replenishment of dusty clouds emerged from accretion disk, resulting in a 
 brightening in the IR emission before the H$\alpha$ enhancement which is not yet observed.  
 On the other hand, \citet[][]{Wang2017} suggested that tidally disrupted clumps within the dust 
 sublimation radius of the torus by central 
 black hole may become bound at smaller radii to serve the source of the BLR gas. 
 {In such a model, the H$\alpha$ anomaly in \src could be induced by a sudden increase in the tidal disruption rate,  
 or in the total mass of tidally disrupted clumps, supplying as addition to the BLR gas clouds. 
 However, the mechanism by which the tidal disruption rate is increased abruptly is not clear in this scenario. 
  In addition, one would ask why the similar behavior of the BEL variations has not been observed in other AGNs 
 for which intense RM monitoring campaigns are available. }


  As we have demonstrated, \src represents one of few AGNs in which 
 a TDE-like energetic transient event was discovered  
  \citep[e.g.,][]{Kankare2017, Blanchard2017, Shu2018, Liu2020}. 
  While the stellar disruption itself may be less affected by the AGN accretion disk, the bound debris stream could 
  collide with the disk, exciting shocks and leading to inflow and considerable energy dissipation in the disk 
  as well as the coronal region responsible for the most AGN X-ray emission \citep{Blanchard2017, Chan2019, Ricci2020}. 
  On the other hand, it has been proposed that when star is disrupted, roughly half the stellar debris is
  unbound, and ejected with an estimated velocity $v_{\rm ej}\sim(2GM_{\rm BH}R_{\star}/R_{\rm p}^2)^{1/2}$ \citep[][]{Evans1989}. 
  For a solar star and with $M_{\rm BH}=7.9\times10^6$\msun~(Section 4.3) for \srcs, 
  {the maximum ejection velocity is of $v_{\rm ej}\sim8\times10^3$ km s$^{-1}$. 
  The ejection of unbound debris may run into the BLR, providing an alternative supply 
  for the BLR gas to explain the H$\alpha$ anomaly. 
  In fact, the geometry and energetics of the ejected material and possible observational consequences of the 
  interaction of this material with the ambient medium surrounding the black hole have been investigated theoretically 
  \citep{Rees1988, Kochanek1994, Khokhlov1996, Guillochon2016}. 
  Given an expected BLR size of $r_{\rm BLR}$$\sim$30 light days from the 5100\AA~luminosity, 
  the timescale for debris to spread out and reach the BLR is $r_{\rm BLR}/v_{\rm ej}=3.1$ years. 
  However, TDE simulations have shown that the average kinetic energy for ejecta is about an order of magnitude lower 
  than expected, resulting in a typical outgoing velocity of $\sim$2500 km s$^{-1}$ \citep{Guillochon2013}. 
  This corresponds to a timescale of $\sim$10 years for debris moving to BLR, which is not 
  inconsistent with the time when the H${\alpha}$ enhancement was observed in \srcs. 
  It is intriguing to note that the changes in the flux of higher-order H$\beta$ line are less than the H$\alpha$ (Figure \ref{halc}), indicating that the gas replenishment alone may not be able to account for the spectral evolution of \srcs, 
  as a larger responsivity of H$\beta$ is expected with decreasing of ionization parameter \citep[][]{Korista2004}. 
  This suggests that in addition to photoionization, collisional excitation might be important for producing H$\alpha$. 
  Better quantitative modeling of the interaction of stellar debris with BLR might be required 
  to explain the anomalous enhancement of H$\alpha$ emission, which is beyond the scope of this paper. 
 }

  Since the anomalous behavior of this kind of BEL variation is rare in AGNs, 
  spectroscopic monitoring campaigns of \src with contemporaneous multi-wavelength 
  photometric observations are encouraged. 
  In addition, observations of similar BEL anomaly in other AGNs from dedicated RM campaigns 
  are important to constrain whether this is unique for \src (e.g., due to the TDE influence) 
  or common phenomenon among AGNs. Such datasets could potentially reveal the underlying physical processes that drive the 
  (de)correlation between continuum and BEL variations, yielding new insights into the formation 
  and evolution of BLR.

\begin{acknowledgements}
 We thank the Swift Acting PI, Brad Cenko, for approving our ToO request to observe \srcs, 
 Matthew J. Graham, Andrew J. Drake for kindly providing the post-DR2 CRTS data for the source, 
  and Pu Du for timely coordinating the Lijiang/YFOSC observations. 
 This research made use of the HEASARC online data archive services, and data products from 
 the Wide-field Infrared Survey Explorer, Catalina Real-time Transient Survey and Zwicky Transient Facility 
 Project. 
 We acknowledge the use of the Hale 200-inch Telescope through the Telescope Access Program (TAP), under 
 the agreement between the National Astronomical Observatories, CAS, and the 
 California Institute of Technology, and the support of the staff of the Lijiang 2.4 m telescope. 
 Funding for the Lijiang telescope has been provided by CAS and the People's Government of Yunnan Province.
 This work is supported by Chinese NSF through grant Nos. 11822301, 12192220, 12192221, 11833007, and U1731104.
\end{acknowledgements}

%
%

 \begin{appendix}

\section{Test on the effects of seeing and slit width on the H$\alpha$ variability}

\begin{figure}[htbp]
\centering
\includegraphics[scale=1.25]{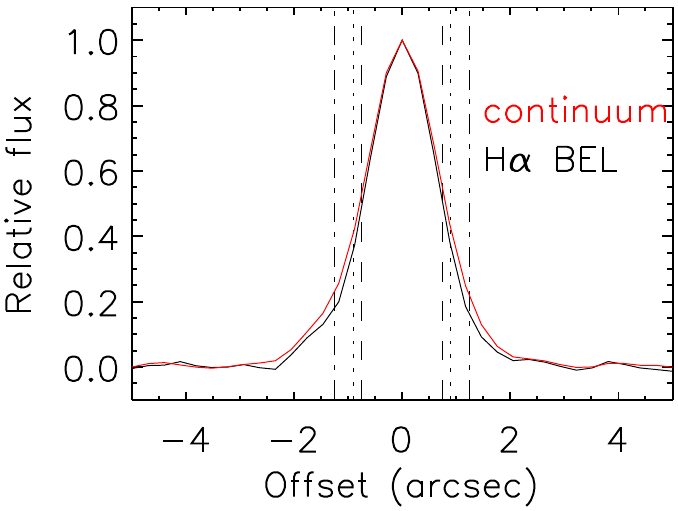}
\caption{
 The brightness profile of H$\alpha$ BEL and underlying continuum along the slit, extracted from the data with the best seeing condition (Feb 2020). 
We also show the slit width of 1.5\arcsec, 1.8\arcsec~and 2.5\arcsec~with dashed, dotted and dash-dotted line, respectively.}
\label{profile}%
\end{figure}

 In this section, we investigate whether the variations in the flux and EW of $H_{\alpha}$ broad emission line (BEL) from 
five long-slit spectroscopic observations can be caused by different 
seeing conditions or slit widths used to extract the spectrum. 
Such an effect could lead to changes in the fraction of emission that falls into the aperture. 
Firstly, we checked whether the spatial brightness profile of the H$\alpha$ BEL and the continuum are different.  
In the case of significant difference in the spatial brightness profile between the two, the measured H$\alpha$ EW may deviate from the actual value.
Since the BLR for an AGN at $z=0.147$ (the redshift of \srcs) is expected to be a point-like source 
for ground-based observations, we can measure the spatial brightness profile of H$\alpha$ BEL along the slit, 
and then compare it to that of the continuum which presumably consists of both AGN and host component. 
To do so, we utilized the data from P200/DBSP observation taken on Feb 2020, as it has the best seeing of 1.5\arcsec. 
The spatial brightness profile of continuum was extracted in two spectral windows around H$\alpha$ with wavelength 
ranges of 6000-6500\AA~and 6640-6900\AA, while the profile of H$\alpha$ BEL was extracted from the continuum-subtracted spectrum 
in the wavelength range of 6505-6615\AA. Note that for extraction of H$\alpha$ BEL, we masked two windows affected by narrow 
emission lines covering narrow wavelength ranges of 6559-6568\AA~and 6582-6588\AA. The results are shown in Figure \ref{profile}. It can be seen that 
the spatial brightness distribution of continuum is slightly extended compared to that of H$\alpha$ BEL, the latter can be considered as a point-like source.

\begin{figure}[htbp]
\centering
\includegraphics[scale=1.2]{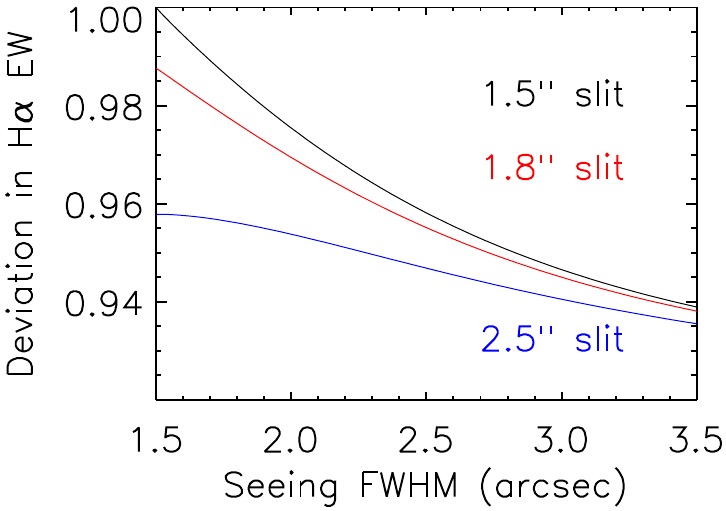}
\caption{
 The deviation of measured H$\alpha$ EW caused by the change in the seeing conditions using simulations, 
for the slit width of 1.5\arcsec~(black), 1.8\arcsec~(red) and 2.5\arcsec~(blue), respectively. }
\label{haew}%
\end{figure}

We demonstrate that the slight difference in the spatial distribution of continuum and H$\alpha$ BEL does 
not cause a significant deviation of the measured EW of H$\alpha$ using the following simulations. In the simulations, 
we assume that the EW of H$\alpha$ remains constant (as the value observed in Feb 2020), and generate fake data that 
would be observed under different seeing conditions. We consider the effect of a worse seeing condition by convolving 
the brightness profile of H$\alpha$ BEL with Gaussian functions. We then measure the fractions of H$\alpha$ BEL and 
continuum that fall into apertures. In Figure \ref{haew}, we show the ratio between the two under different seeing conditions 
for a slit width of 1.5\arcsec, 1.8\arcsec~and 2.5\arcsec, respectively. The ratios are normalized to the value measured in Feb 2020. 
This normalized ratio represents the deviation of the measured EW of H$\alpha$ caused by changes in different seeing conditions 
and slit widths. The simulations show that the deviation is at most at a level of 7\% for the seeing between 1.5\arcsec~and 3.3\arcsec, 
and the slit width between 1.5\arcsec~and 2.5\arcsec. 
Note that for the other four spectroscopic observations with worse seeing conditions, the deviation is expected smaller.
Since the observed variations in the EW of H$\alpha$ have a much larger 
amplitude than this value (up to 33\%), we conclude that the H$\alpha$ variability is intrinsic 
and cannot be explained by the variations in either seeing or slit width. 

\section{UV to optical SED fittings}

\begin{figure}[htbp]
 \centering
 \includegraphics[scale=0.33]{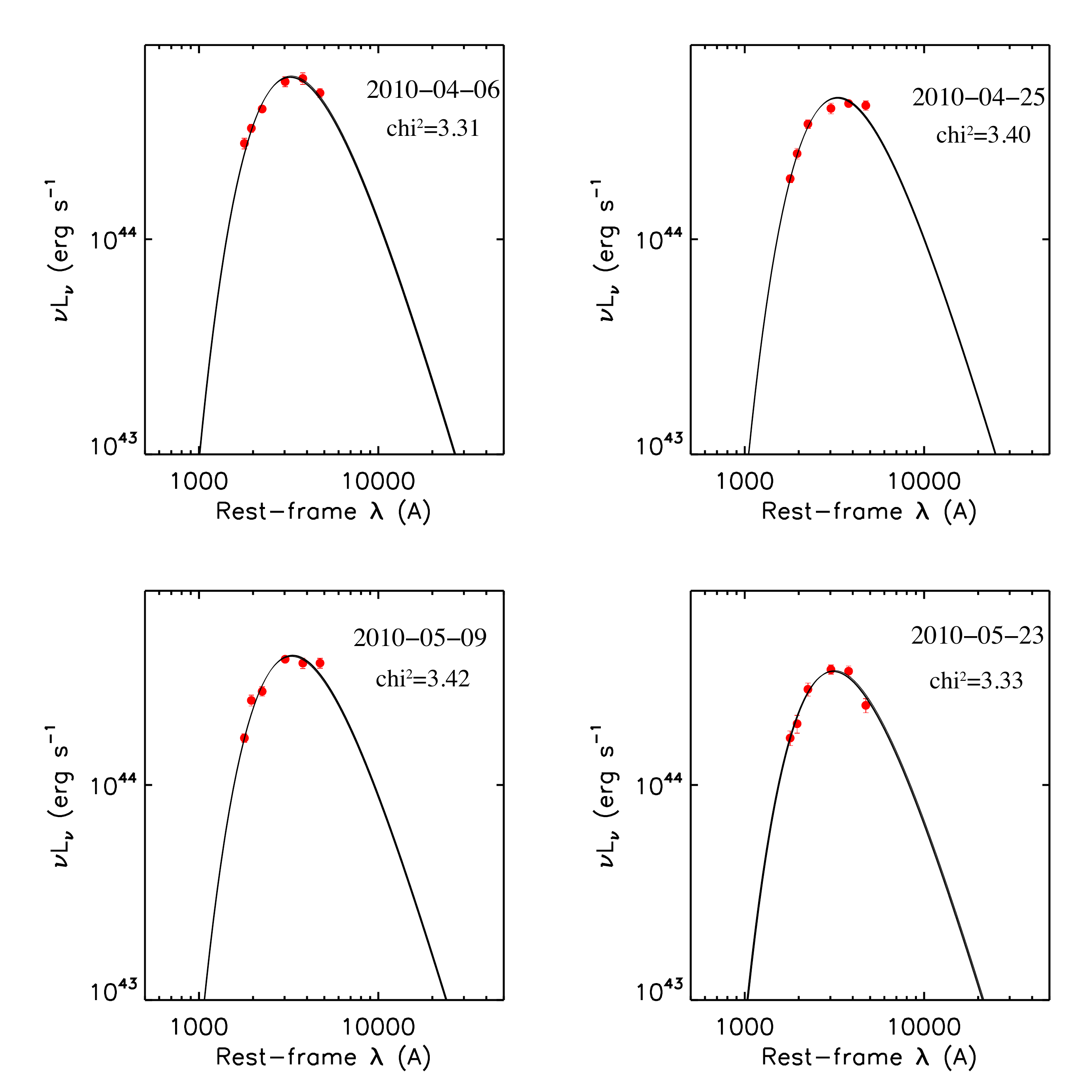}
 \caption{
UV to optical SED obtained from the \swift UVOT observations during the period of the outburst, with the best-fit blackbody model for each epoch. 
 }
 \label{bbody}%
 \end{figure}

\begin{figure}[htbp]
 \centering
 \includegraphics[scale=0.4]{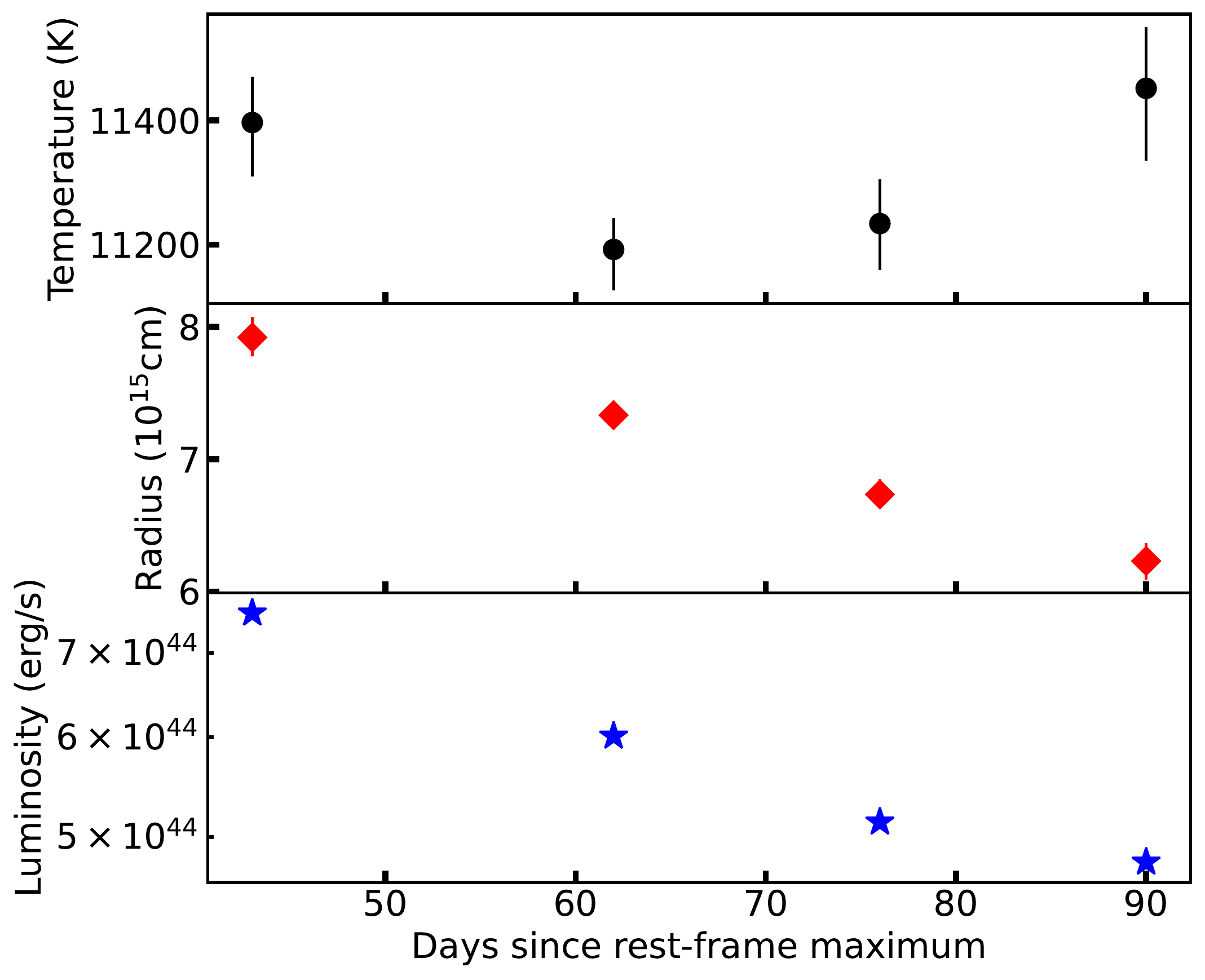}
 \caption{
 The evolution of UV/optical emission of \src at different epochs. From top to bottom panel, we 
 show the evolution of blackbody luminosity, temperature and radius, respectively, which are derived 
 from blackbody fittings to the \swift UVOT data. 
 }
 \label{bbodyevo}%
 \end{figure}

{We fitted a blackbody model ($B_{\nu}=\frac{2h\nu^3}{c^2}\frac{1}{e^{h\nu/kT}-1}$) to the host subtracted, 
extinction-corrected photometric data from the \swift UVOT observations during the period of optical outburst, to put constrains on the luminosity, 
temperature and radius evolution of UV and optical emission.  
    Uncertainties on the above parameters were derived using Monte Carlo simulations, in which the observed 
    fluxes were randomly perturbed with amplitude by assuming Gaussian noise according to 
    the photometric errors. This procedure was repeated 1000 times.  
    The error bars on each parameter were then derived from the 16th 
    and 84th percentiles of the distribution of the corresponding values 
    obtained in the simulations. 
    The UV to optical SED for different epochs, with its best-fitting blackbody model, 
    are shown in Figure \ref{bbody}. 
    Using the best-fit temperature and rest-frame monochromatic 
       UV luminosity at each epoch, we estimated the blackbody radius from 
       $L_\nu=\pi B_\nu\times4\pi R_{\rm BB}^2$ and took blackbody luminosity 
       from $L_{\rm BB}=\sigma T^4 \times4\pi R_{\rm BB}^2$ as the integrated 
       luminosity of UV and optical emission.  The evolution of blackbody luminosity, 
       temperature and radius are presented in the Figure \ref{bbodyevo}.
    }

\end{appendix}

\clearpage

\end{document}